\documentstyle[12pt, epsf]{article}

\advance \topmargin by -\headheight
\advance \topmargin by -\headsep

\evensidemargin \oddsidemargin
 \def\mod{{\hbox{\rm mod}}}

\def\ie{{\em i.e.}}
\def\ie{\hbox{\it i.e.}}

\def\CC{{\mathchoice
{\rm C\mkern-8mu\vrule height1.45ex depth-.05ex
width.05em\mkern9mu\kern-.05em}
{\rm C\mkern-8mu\vrule height1.45ex depth-.05ex
width.05em\mkern9mu\kern-.05em}
{\rm C\mkern-8mu\vrule height1ex depth-.07ex
width.035em\mkern9mu\kern-.035em}
{\rm C\mkern-8mu\vrule height.65ex depth-.1ex
width.025em\mkern8mu\kern-.025em}}}

\def\RR{{\rm I\kern-1.6pt {\rm R}}}

\def\ZZ{{\rm Z}\kern-3.8pt {\rm Z} \kern2pt}
\def\IB{\relax{\rm I\kern-.18em B}}
\def\ID{\relax{\rm I\kern-.18em D}}
\def\II{\relax{\rm I\kern-.18em I}}
\def\IP{\relax{\rm I\kern-.18em P}}

\def\np{Nucl. Phys.}
\def\pl{Phys. Lett.}
\def\prl{Phys. Rev. Lett.}
\def\pr{Phys. Rev.}

\def\ijmp{Int. J. Mod. Phys.}

\def\jhep{J. High Energy Phys.}

\newcommand{\beq}{\begin{equation}}
\newcommand{\eeq}{\end{equation}}
\newcommand{\rc}{\nonumber\\}
\newcommand{\bear}{\begin{eqnarray}}
\newcommand{\eear}{\end{eqnarray}}

\def\to{\rightarrow}

\def\to{\rightarrow}

\newfont{\namefont}{cmr10}
\newfont{\addfont}{cmti7 scaled 1440}
\newfont{\boldmathfont}{cmbx10}
\newfont{\headfontb}{cmbx10 scaled 1728}
\renewcommand{\theequation}{{\rm\thesection.\arabic{equation}}}
\begin{document}
\begin{titlepage}

\begin{center} \Large \bf Flavoring  the gravity dual of  ${\cal
N}=1$ Yang-Mills\\ with probes

\end{center}

\vskip 0.3truein
\begin{center}
Carlos N\'u\~nez${}^{\,\dagger}$
\footnote{nunez@lns.mit.edu},
\'Angel Paredes${}^{\,*}$
\footnote{angel@fpaxp1.usc.es}
and
Alfonso V. Ramallo${}^{\,*}$
\footnote{alfonso@fpaxp1.usc.es}

\vspace{0.3in}
${}^{\dagger\,}$Center for Theoretical Physics, Massachusetts Institute of Technology \\
Cambridge, MA 02139, USA

\vspace{0.3in}

${}^{\,*}$Departamento de F\'\i sica de Part\'\i culas, Universidad de
Santiago de Compostela \\
E-15782 Santiago de Compostela, Spain
\vspace{0.3in}

\end{center}
\vskip 1truein

\begin{center}
\bf ABSTRACT
\end{center}
We study two related problems in the context of a supergravity dual to  ${\cal N}=1$
SYM. One of the problems is finding kappa symmetric D5-brane probes in this 
particular background. The other is the use
of these probes to add flavors to the gauge theory.
We find a rich and mathematically appealing structure of the supersymmetric 
embeddings of a D5-brane probe in this background. Besides, we compute the mass spectrum of
the  low energy excitations of ${\cal N}=1$ SQCD (mesons) and match our results with some
field  theory aspects known from the study of supersymmetric gauge theories with
a small number of flavors.

\vskip2.6truecm
\leftline{US-FT-3/03 }
\leftline{MIT-CPT/3441}
\leftline{hep-th/0311201 \hfill November 2003}
\smallskip
\end{titlepage}
\setcounter{footnote}{0}


\setcounter{equation}{0}
\section{Introduction}
\medskip
The gauge/string correspondence, an old proposal due to 't Hooft \cite{Hooft},  is now
well understood in the context of maximally supersymmetric super Yang-Mills (SYM)
theories. Indeed, the so-called AdS/CFT correspondence is a conjectured equivalence
between type IIB string theory on $AdS_5\times S^5$ and ${\cal N}=4$ SYM theory \cite{jm}.
In the large  't Hooft coupling limit, the  ${\cal N}=4$ SYM theory is dual to the type
IIB supergravity background corresponding to the near-horizon geometry of a stack of
parallel D3-branes, whose metric is precisely that of the $AdS_5\times S^5$ space.
There are nowadays a lot of non-trivial tests of this duality (for a review see
\cite{magoo}). 

The extension of the gauge/string correspondence to theories with less supersymmetries
is obviously of great interest. A possible way to obtain supergravity duals of SYM
theories with reduced supersymmetry is to consider branes wrapping supersymmetric
cycles of Calabi-Yau manifolds \cite{MNfirst}. In order to preserve some supersymmetry the
normal bundle of the cycle within the Calabi-Yau space has to be twisted \cite{bvs}.
Gauged supergravities in lower dimensions provide the most natural framework to implement
this twisting. In these theories the gauge field can be used to fiber the cycle in
which the brane is wrapped in such a way that some supersymmetries are preserved.

In this paper we will restrict ourselves to the case of the supergravity dual of 
${\cal N}=1$ SYM. This background, which corresponds to a fivebrane wrapping a
two-cycle, was obtained in ref. \cite{MN} from the solution found
in ref. \cite{CV} representing non-abelian magnetic monopoles in four dimensions. The
geometry of this background is smooth and leads to confinement and chiral symmetry
breaking. Actually, if only the abelian part of the vector field of seven dimensional
gauged supergravity is excited, one obtains a geometry which is singular at the origin
and coincides with the smooth one at large distances, \ie\ in the UV. Therefore, the
singularity at the origin is resolved by making the gauge field non-abelian, in complete
analogy to what happens with the resolution of the Dirac string by the 't Hooft-Polyakov
monopole. Moreover, as argued in ref. \cite{Apreda}, the same mechanism that
de-singularizes the supergravity solution also gives rise to gaugino condensation. Based
on this observation, the NSVZ beta function can be reproduced at leading order \cite{DV,
BertMer,Merlatti}. Other aspects of this supergravity dual have been studied in ref.
\cite{SeveralMN}(for a review see \cite{MerReview}).

Most of the analysis carried out with the background of \cite{MN} do
not incorporate quarks in the fundamental representation which, in a string theory setup,
correspond to open strings. In order to introduce an open string sector in a
supergravity dual it is quite natural to add D-brane probes and see whether one can
extract some information about the quark dynamics. As usual, if the number of brane
probes is much smaller than those of the background, one can assume that there is no
backreaction of the probe in the bulk geometry. In this paper we follow this approach and
we will probe with D5-branes the supergravity dual of  ${\cal N}=1$ SYM. The main
technique to determine the supersymmetric brane probe configurations is kappa symmetry
\cite{swedes}, which
tells us that, if $\epsilon$ is a Killing spinor of the background, only those embeddings
for which a certain matrix $\Gamma_{\kappa}$ satisfies
\beq
\Gamma_{\kappa}\,\epsilon\,=\,\epsilon\,\,.
\label{kappaprojection}
\eeq
preserve the supersymmetry of the background \cite{bbs}. The matrix $\Gamma_{\kappa}$ 
depends on the metric induced on the worldvolume of the brane. Therefore, if the Killing
spinors $\epsilon$ are known, we can regard  (\ref{kappaprojection}) as an equation for
the embedding of the brane.

The starting point in our program will be  the determination of the Killing spinors of the
background. It turns out that a simple expression for these spinors can be obtained if
one considers a frame inspired by the uplifting of the metric from gauged seven
dimensional supergravity. The realization of the topological twist in this case is
similar to the one introduced in ref. \cite{Twist}, 
which generalizes that of ref. \cite{gaugedsugra},
to obtain manifolds of $G_2$
holonomy and the deformed and resolved conifold from gauged eight-dimensional
supergravity. The resulting Killing spinors are characterized by a series of projections
and all we have to do is to find those configurations for which the kappa symmetry
condition (\ref{kappaprojection})  follows from the projections satisfied by the Killing
spinors.  

The probes we are going to consider are D5-branes wrapped on a two-dimensional
submanifold. We will be able to find some differential equations for the embedding which
are, in general, quite complicated to solve. The first obvious configuration one should
look at is that of a fivebrane wrapped at a fixed distance from the origin. In this case
the equations simplify drastically and we will be able to prove a no-go theorem which
states that, unless we place the brane  at an infinite distance from the origin, the
probe breaks supersymmetry. This result is consistent with the fact that these ${\cal
N}=1$ theories do not have a moduli space. In this analysis we will make contact with the
two-cycle considered in ref. \cite{BertMer} and show that it preserves supersymmetry at an
asymptotically large distance from the origin. 

Guided by the negative result obtained when trying to wrap the D5-brane at constant
distance, we will allow this distance to vary within the two-submanifold of the embedding.
To simplify the equations that determine the embeddings, we first consider the singular
version of the background, in which the vector field of the seven dimensional gauged
supergravity is abelian. This geometry coincides with the non-singular one, in which the
vector field is non-abelian, at large distances from the origin. By choosing an
appropriate set of variables we will be able to write the differential equations for the
embedding as two pairs of Cauchy-Riemann equations which are straightforward to integrate
in general. Among all possible solutions we will concentrate on some of them characterized
by integers, which can be interpreted as winding numbers. Generically these solutions have
spikes, in which the probe is at infinite distance from the origin and, thus, they
correspond to fivebranes wrapping a non compact submanifold. Moreover, these
configurations are worldvolume solitons and we will verify that they saturate an energy
bound
\cite{GGT}.

With the insight gained by the analysis of the worldvolume solitons in the abelian
background we will consider the equations for the embeddings in the non-abelian
background. In principle any solution for the smooth geometry must coincide in the UV
with one of the configurations found for the singular metric. This observation will allow
us to formulate an ansatz to solve the complicated equations arising from kappa symmetry.
Actually, in some cases,  we will be able to find analytical solutions for the embeddings,
which behave as those found for the singular metric at large distance from the origin and
also saturate an energy bound, which ensures their stability.

One of our motivations to study brane probes is to use these results to explore the quark
sector of the gauge/gravity duality. Actually, it was proposed in refs. \cite{KK, KKW}
that one can add flavor to this correspondence by considering spacetime filling branes
and looking at their fluctuations. In ref. \cite{KMMW} this program has been made
explicit for the $AdS_5\times S^5$ geometry of a stack of D3-branes and a D7-brane
probe. When the D3-branes of the background and the D7-brane of the probe are separated,
the fundamental matter arising from the strings stretched between them becomes massive and
a  discrete spectrum  of mesons for an ${\cal N}=2$ SYM with a matter hypermultiplet can
be obtained analytically from the fluctuations of a D7-brane probe. In ref. \cite{Sonnen}
a similar analysis was performed for the ${\cal N}=1$ Klebanov-Strassler background
\cite{KS}, while in refs. \cite{Johana,KMMW-two} the meson spectrum for some
non-supersymmetric backgrounds was found (for recent related work see refs.
\cite{IL,Ouyang}).

It was suggested in ref. \cite{WangHu} that one possible way to add flavor to the  
${\cal N}=1$ SYM background is by considering supersymmetric embeddings of
D5-branes which wrap a two-dimensional submanifold and are spacetime filling. Some of the
configurations we will find in our kappa symmetry analysis have the right ingredients to
be used as flavor branes. They are supersymmetric by construction, extend infinitely and
have some parameter which determines the minimal distance between the brane probe and the
origin. This distance should be interpreted as the mass scale of the quarks. Moreover,
these brane probes capture geometrically the pattern of R-symmetry breaking of SQCD with
few flavors
\cite{Affleck:1983mk}. Consequently, we will study the quadratic fluctuations
around the static probe configurations found by integrating the kappa symmetry
equations. We will verify that these fluctuations decay exponentially at large
distances. However, we will not be able to define a normalizability condition which
could give rise to a discrete spectrum. The reason for this is the exponential blow up
of the dilaton at large distances. Actually, this same difficulty was found in ref.
\cite{Pons} in the study of the glueball spectrum for this background. As proposed in
ref. \cite{Pons}, we shall introduce a cut-off and impose boundary conditions which
ensure that the fluctuation takes place in a region in which the supergravity
approximation remains valid. The resulting spectrum is discrete and, by using numerical
methods, we will be able to determine its form. 

The organization of this paper is the following. In section 2 we introduce the
supergravity dual of  ${\cal N}=1$ SYM. The Killing spinors for this background are
obtained in appendix A, where we also obtain those corresponding to the background of
refs. \cite{Martin, MaldaNas, CVdos}, which represents D5-branes wrapped on a three-cycle.
In section 3 we obtain the kappa symmetry equations which determine the supersymmetric
embeddings. In section 4 we obtain the no-go theorem for branes wrapped at fixed
distance. In section 5 the kappa symmetry equations for the abelian background are
integrated in general and some of the particular solutions are studied in detail. Section
6 deals with the integration of the equations for the supersymmetric embeddings in the
full non-abelian background. The spectrum of the quadratic fluctuations is analysed in
section 7. The asymptotic form of these fluctuations is obtained in appendix B. Finally,
in section 8 we summarize our results, draw some conclusions and discuss some lines of
future work.

\subsection{Reader's guide}

Given that this is a long paper, we feel it would be useful to include here a
``roadmap" to help the reader to quickly find the results of his/her particular interest. Those
readers interested in the supersymmetry preserved by the background and in the application of
kappa symmetry to find compact and non compact embeddings in the geometry dual to 
${\cal N}=1$ SYM, should pay special attention to sections 2-6 and appendix A. Readers more
interested in the gravity version of the addition of flavors to ${\cal N}=1$ SYM should
take for granted section 3 and look at the solutions exhibited in eqs. (\ref{abspikes}),
(\ref{BM}) and (\ref{noabflavor}), which are what we called ``abelian and non-abelian
unit-winding solutions". Then, they should go straight to section 7 and take into account
the results of appendix B.

\setcounter{equation}{0}
\section{The supergravity dual of  ${\cal N}=1$ Yang-Mills}
\medskip
The supergravity solution we will be dealing with corresponds to a stack of $N$ D5-branes
wrapped on a two-cycle. It can be obtained \cite{MN,CV} by considering seven dimensional
gauged supergravity, which is a consistent truncation of ten dimensional supergravity on
a three-sphere. To get this background one starts with an ansatz for the seven
dimensional metric which has a term corresponding to the metric of a  two-sphere and
looks for a supersymmetric solution of the equations of motion. After uplifting to ten
dimensions one ends up with a solution of type IIB supergravity which preserves four
supersymmetries. The ten dimensional metric in Einstein frame is:
\beq
ds^2_{10}\,=\,e^{{\phi\over 2}}\,\,\Big[\,
dx^2_{1,3}\,+\,e^{2h}\,\big(\,d\theta^2+\sin^2\theta d\varphi^2\,\big)\,+\,
dr^2\,+\,{1\over 4}\,(w^i-A^i)^2\,\Big]\,\,,
\label{metric}
\eeq
where $\phi$ is the dilaton, the unwrapped coordinates $x^{\mu}$  have been
rescaled and all distances are measured in units of $Ng_s\alpha'$. The angles
$\theta\in [0,\pi]$ and 
$\varphi\in [0,2\pi)$ parametrize the two-sphere of gauged seven dimensional
supergravity. This sphere is fibered in the ten dimensional metric by the one-forms 
$A^i$ $(i=1,2,3)$, which are the components of the non-abelian gauge vector field of the
seven dimensional supergravity. Their expression can be written in terms of a function 
$a(r)$ and the angles $(\theta,\varphi)$ as follows:
\beq
A^1\,=\,-a(r) d\theta\,,
\,\,\,\,\,\,\,\,\,
A^2\,=\,a(r) \sin\theta d\varphi\,,
\,\,\,\,\,\,\,\,\,
A^3\,=\,- \cos\theta d\varphi\,.
\label{oneform}
\eeq
The   $w^i\,$'s appearing in eq. (\ref{metric}) are the $su(2)$ left-invariant one-forms,
satisfying 
$dw^i=-{1\over 2}\,\epsilon_{ijk}\,w^j\wedge w^k$, which parametrize the compactification
three-sphere and can be represented in
terms of three angles $\tilde\varphi$, $\tilde\theta$ and $\psi$:
\bear
w^1&=& \cos\psi d\tilde\theta\,+\,\sin\psi\sin\tilde\theta
d\tilde\varphi\,\,,\rc\rc
w^2&=&-\sin\psi d\tilde\theta\,+\,\cos\psi\sin\tilde\theta 
d\tilde\varphi\,\,,\rc\rc
w^3&=&d\psi\,+\,\cos\tilde\theta d\tilde\varphi\,\,.
\eear
The three angles $\tilde\varphi$, $\tilde\theta$ and $\psi$ take values in the
rank $0\le\tilde\varphi< 2\pi$, $0\le\tilde\theta\le\pi$ and
$0\le\psi< 4\pi$. For a metric ansatz such as the one written in 
(\ref{metric}) one  obtains a supersymmetric solution when
the functions $a(r)$, $h(r)$ and the dilaton $\phi$ are:
\bear
a(r)&=&{2r\over \sinh 2r}\,\,,\rc\rc
e^{2h}&=&r\coth 2r\,-\,{r^2\over \sinh^2 2r}\,-\,
{1\over 4}\,\,,\rc
e^{-2\phi}&=&e^{-2\phi_0}{2e^h\over \sinh 2r}\,\,,
\label{MNsol}
\eear
where $\phi_0$ is the value of the dilaton at $r=0$. Near the origin $r=0$ the function 
$e^{2h}$ behaves as $e^{2h}\sim r^2$ and the metric is non-singular. The solution of the
type IIB supergravity includes a
Ramond-Ramond three-form $F_{(3)}$ given by
\beq
F_{(3)}\,=\,-{1\over 4}\,\big(\,w^1-A^1\,\big)\wedge 
\big(\,w^2-A^2\,\big)\wedge \big(\,w^3-A^3\,\big)\,+\,{1\over 4}\,\,
\sum_a\,F^a\wedge \big(\,w^a-A^a\,\big)\,\,,
\label{RRthreeform}
\eeq
where $F^a$ is the field strength of the su(2) gauge field $A^a$, defined as:
\beq
F^a\,=\,dA^a\,+\,{1\over 2}\epsilon_{abc}\,A^b\wedge A^c\,\,.
\label{fieldstrenght}
\eeq
The different components of $F^a$ are:

\beq
F^1\,=\,-a'\,dr\wedge d\theta\,\,,
\,\,\,\,\,\,\,\,\,\,
F^2\,=\,a'\sin\theta dr\wedge d\varphi\,\,,
\,\,\,\,\,\,\,\,\,\,
F^3\,=\,(\,1-a^2\,)\,\sin\theta d\theta\wedge d\varphi\,\,,
\eeq
where the prime denotes derivative with respect to $r$. 
Since $dF_{(3)}=0$, one can represent $F_{(3)}$ in terms of a two-form potential 
$C_{(2)}$ as $F_{(3)}\,=\,dC_{(2)}$. Actually, it is not difficult to verify that 
$C_{(2)}$ can be taken as:
\bear
C_{(2)}&=&{1\over 4}\,\Big[\,\psi\,(\,\sin\theta d\theta\wedge d\varphi\,-\,
\sin\tilde\theta d\tilde\theta\wedge d\tilde\varphi\,)
\,-\,\cos\theta\cos\tilde\theta d\varphi\wedge d\tilde\varphi\,-\rc\rc
&&-a\,(\,d\theta\wedge w^1\,-\,\sin\theta d\varphi\wedge w^2\,)\,\Big]\,\,.
\eear
Moreover, the equation of motion of $F_{(3)}$ in the Einstein frame is
$d\Big(\,e^{\phi}\,{}^*F_{(3)}\,\Big)=0$, where $*$ denotes Hodge duality. Therefore
it follows that, at least locally, one must have
\beq
e^{\phi}\,{}^*F_{(3)}\,=\,d C_{(6)}\,\,,
\eeq
with $C_{(6)}$ being a six-form potential. It is readily checked that 
$C_{(6)}$ can be taken as:
\beq
C_{(6)}\,=\,dx^0\wedge dx^1\wedge dx^2\wedge dx^3\wedge
{\cal C}\,\,,
\eeq
where ${\cal C}$ is the following two-form:
\bear
{\cal C}&=&-{e^{2\phi}\over 8}\,\,\Big[\,
\Big(\,(\,a^2-1\,)a^2\,e^{-2h}\,-\,16\,e^{2h}\,\Big)\,\cos\theta
d\varphi\wedge dr
\,-\,(\,a^2-1\,)\,e^{-2h}\,w^3\wedge dr\,+\rc\rc
&&+\,a'\,\Big(\,\sin\theta d\varphi\wedge w^1\,+\,d\theta\wedge
w^2\,\Big)\,\Big]\,\,.
\eear

The Killing spinors $\epsilon$ of the above background are worked out in Appendix A. They
can be obtained by requiring the vanishing of the supersymmetry variations of the
fermionic fields of type IIB supergravity. This requirement leads to a system of
first-order BPS differential equations (eq. 
(\ref{differentialeqs})) for the functions $\phi$, $h$ and $a$  of the ansatz written in
eqs. (\ref{metric}) and (\ref{RRthreeform}). It can be easily checked that the functions
of eq. (\ref{MNsol}) satisfy the system (\ref{differentialeqs}). Moreover, it follows
from the analysis of appendix A that the Killing spinors $\epsilon$ 
are characterized by the following set of projections:
\bear
&&\Gamma_{x^0\cdots x^3}\,\big(\,
\cos\alpha\,\Gamma_{12}\,+\,\sin\alpha\,\Gamma_1\hat\Gamma_2\,)\,
\epsilon\,=\,\epsilon\,\,,\rc\rc
&&\Gamma_{12}\,\epsilon\,=\,\hat\Gamma_{12}\,\epsilon\,\,,\rc\rc
&&\epsilon\,=\,i\epsilon^*\,\,,
\label{fullprojection}
\eear
where the $\Gamma$-matrices refer to the frame (\ref{frame}) and the explicit
expression of the angle $\alpha$, which depends on the radial coordinate $r$, is given in
eqs. (\ref{alpha}) and  (\ref{alphaexplicit}) (see eqs. (\ref{projectionone}),
(\ref{projectiontwo}) and (\ref{alphaproj})). It is interesting to write here the UV and
IR limits of $\alpha$, namely
\beq
\lim_{r\rightarrow \infty}\,\alpha\,=\,0\,\,,
\,\,\,\,\,\,\,\,\,\,\,\,\,\,\,\,\,\,\,\,
\lim_{r\rightarrow 0}\,\alpha\,=\,-{\pi\over 2}\,\,.
\eeq

The BPS equations (\ref{differentialeqs}) also admit a solution in which the function
$a(r)$ vanishes, \ie\ in which the one-form $A^i$ has only one non-vanishing
component, namely $A^{3}$. We will refer to this solution as the abelian ${\cal N}=1$
background. Its explicit form can be easily obtained by taking the
$r\rightarrow\infty$ limit of the functions given in eq. (\ref{MNsol}). Notice that,
indeed $a(r)\rightarrow 0$ as $r\rightarrow\infty$ in  eq. (\ref{MNsol}). 
Neglecting exponentially suppressed terms, one gets:
\beq
e^{2h}\,=\,r\,-\,{1\over 4}\,\,,
\,\,\,\,\,\,\,\,\,\,\,\,\,\,\,\,\,\,(a=0)\,\,,
\eeq
while $\phi$ can be obtained from the last equation in (\ref{MNsol}). The metric
of the abelian background is singular at $r=1/4$ (the position of the singularity can
be moved to $r=0$ by a redefinition of the radial coordinate). This IR singularity of
the abelian background is removed in the non-abelian metric by switching on the $A^1,
A^2$ components of the one-form (\ref{oneform}). Moreover, when $a=0$, the angle
$\alpha$ appearing in the expression of the Killing spinors (and in the projection 
(\ref{fullprojection})) is zero, as  follows from eq. (\ref{alpha}).

\setcounter{equation}{0}
\section{Kappa symmetry}
\medskip
As mentioned in the introduction, 
the kappa symmetry condition for a supersymmetric embedding of a D5-brane probe is
$\Gamma_{\kappa}\,\epsilon\,=\,\epsilon$ (see eq. (\ref{kappaprojection})), 
where $\epsilon$ is a Killing spinor of the background. For $\epsilon$ such that
$\epsilon=i\epsilon^*$  and when there is no worldvolume gauge field, one has:
\beq
\Gamma_{\kappa}\,=\,{1\over 6!}\,\,{1\over \sqrt{-g}}\,\,
\epsilon^{m_1\cdots m_6}\,\,\gamma_{m_1\cdots m_6}\,\,,
\eeq
where $g$ is the determinant of the induced metric $g_{mn}$ on the worldvolume 
\beq
g_{mn}\,=\,\partial_m X^{\mu}\,\partial_n X^{\nu}\,G_{\mu\nu}\,\,,
\eeq
with $G_{\mu\nu}$ being the ten-dimensional metric
and
$\gamma_{m_1\cdots m_6}$ are antisymmetrized products of worldvolume Dirac matrices 
$\gamma_{m}$, defined as:
\beq
\gamma_{m}\,=\,\partial_{m}X^{\mu}\,E_{\mu}^{\underline{\nu}}\,
\Gamma_{\underline{\nu}}\,\,.
\eeq
The vierbeins $E_{\mu}^{\underline{\nu}}$ are the coefficients which relate the
one-forms $e^{\underline{\nu}}$ of the frame and the differentials of the
coordinates, \ie\ $e^{\underline{\nu}}=E_{\mu}^{\underline{\nu}}dX^{\mu}$. 
Let us take as worldvolume coordinates $(x^0, \cdots, x^3, \theta,\varphi)$.
Then, for an embedding with $\tilde\theta=\tilde\theta(\theta,\varphi)$, 
$\tilde\varphi=\tilde\varphi(\theta,\varphi)$, 
$\psi=\psi(\theta,\varphi)$ and $r=r(\theta, \varphi)$, the kappa symmetry matrix
$\Gamma_{\kappa}$ takes the form:
\beq
\Gamma_{\kappa}\,=\,{e^{\phi}\over \sqrt{-g}}\,\Gamma_{x^0\cdots x^3}\,
\gamma_{\theta\varphi}\,\,,
\eeq	
with $\gamma_{\theta\varphi}$ being the antisymmetrized product of the two induced
matrices
$\gamma_{\theta}$ and $\gamma_{\varphi}$, which can be written as:
\bear
e^{-{\phi\over 4}}\,\gamma_{\theta}&=&
e^h\,\Gamma_1\,+\,(V_{1\theta}+{a\over 2}\,)\,\hat\Gamma_1\,+\,
V_{2\theta}\,\hat\Gamma_2\,+\,V_{3\theta}\,\hat\Gamma_3\,+\,
\partial_{\theta} r\Gamma_{r}\,\,,\rc\rc
{e^{-{\phi\over 4}}\over \sin\theta}\,\gamma_{\varphi}&=&
e^h\,\Gamma_2\,+\,
V_{1\varphi}\,\hat\Gamma_1\,+\,(V_{2\varphi}\,-{a\over 2}\,)\,\hat\Gamma_2
\,+\,V_{3\varphi}\,\hat\Gamma_3\,+\,
{\partial_{\varphi} r\over \sin\theta}\,
\Gamma_r
\,\,,
\label{gthetaphi}
\eear
where the $V$'s can be obtained by computing the pullback on the worldvolume of the left
invariant one-forms
$w^i$, and are given by:
\bear
V_{1\theta}&=&{1\over 2}\,\cos\psi\,\partial_{\theta}\,\tilde\theta\,+\,
{1\over 2}\sin\psi\sin\tilde\theta\,\partial_{\theta}\,\tilde\varphi\,\,,\rc\rc
\sin\theta\,V_{1\varphi}&=&
{1\over 2}\,\cos\psi\,\partial_{\varphi}\,\tilde\theta\,+\,
{1\over 2}\sin\psi\sin\tilde\theta\,\partial_{\varphi}\,\tilde\varphi\,\,,\rc\rc
V_{2\theta}&=&-{1\over 2}\,\sin\psi\,\partial_{\theta}\,\tilde\theta\,+\,
{1\over 2}\cos\psi\sin\tilde\theta\,\partial_{\theta}\,\tilde\varphi\,\,,\rc\rc
\sin\theta\,V_{2\varphi}&=&-
{1\over 2}\,\sin\psi\,\partial_{\varphi}\,\tilde\theta\,+\,
{1\over 2}\cos\psi\sin\tilde\theta\,\partial_{\varphi}\,\tilde\varphi\,\,,\rc\rc
V_{3\theta}&=&{1\over 2}\,\partial_{\theta}\psi\,+\,{1\over 2}\cos\tilde\theta\,
\partial_{\theta}\,\tilde\varphi\,\,,\rc\rc
\sin\theta\,V_{3\varphi}&=&
{1\over 2}\,\partial_{\varphi}\psi\,+\,{1\over 2}\cos\tilde\theta\,
\partial_{\varphi}\,\tilde\varphi\,+\,{1\over 2}\,\cos\theta\,\,.
\label{Vs}
\eear
By using the projections (\ref{projectionone}) and (\ref{radial}) one can compute the
action of $\gamma_{\theta\varphi}$ on the Killing spinor $\epsilon$. It is clear that
one arrives at an expression of the type:
\bear
{e^{-{\phi\over 2}}\over \sin\theta}\,\,
\gamma_{\theta\varphi}\,\epsilon\,&=&\,\big[\,
c_{12}\,\Gamma_{12}\,+\,c_{1\hat 2}\,\Gamma_{1}\hat\Gamma_{2}\,+\,
c_{1\hat 1}\,\Gamma_{1}\hat\Gamma_{1}\,+\,
c_{1\hat 3}\,\Gamma_{1}\hat\Gamma_{3}\,+\,
\rc\rc
&&+\,c_{\hat 1\hat 3}\,\hat\Gamma_{13}\,
+\,c_{\hat 2\hat 3}\,\hat\Gamma_{23}\,
+\, c_{2\hat 3}\,\Gamma_{2}\hat\Gamma_{3}\,\big]\,\epsilon\,\,,
\label{ces}
\eear
where the $c$'s are coefficients that can be explicitly computed. By using eq.
(\ref{ces}) we can obtain the action of $\Gamma_{\kappa}$ on $\epsilon$ and we can use
this result to write the kappa symmetry projection $\Gamma_{\kappa}\epsilon=\epsilon$.
Actually, eq. (\ref{kappaprojection}) is automatically satisfied if it reduces to the
first equation in eq. (\ref{fullprojection}). If we want this to happen,
all terms except the ones containing $\Gamma_{12}\,\epsilon$ and 
$\Gamma_{1}\,\hat\Gamma_{2}\,\epsilon$ on the right-hand side of eq. (\ref{ces}) should
vanish. Then, we should require
\beq
c_{1\hat 1}\,=\,c_{1\hat 3}\,=\,c_{\hat 1\hat 3}\,=\,c_{\hat 2\hat 3}\,=\,
c_{2\hat 3}\,=\,0\,\,.
\label{cesnull}
\eeq
By using the explicit expressions of the $c$'s one can obtain from eq. (\ref{cesnull})
five conditions that our supersymmetric embeddings must necessarily satisfy. These
conditions are:
\bear
&&e^{h}\,(V_{1\varphi}\,+\,V_{2\theta}\,)=0\,\,,\label{kappauno}\\\rc
&&e^h\,(\,V_{3\varphi}\,+\,\cos\alpha\partial_{\theta}r\,)\,
+\,(\,V_{2\varphi}\,-\,{a\over 2}\,)\,\sin\alpha\,\partial_{\theta}r
\,-\,V_{2\theta}\,\sin\alpha\,\,{\partial_{\varphi} r\over \sin\theta}
\,=\, 0\,\,,\label{kappados}\\\rc
&&(\,V_{1\theta}\,+\,{a\over 2}\,)\,V_{3\varphi}\,-\,
V_{3\theta}\,V_{1\varphi}\,-\,e^{h}\sin\alpha\,\partial_{\theta}r\,+\,\rc\rc
&&\,\,\,\,\,\,\,\,\,\,\,\,\,\,\,\,\,\,\,\,\,\,\,
+\,(\,V_{2\varphi}\,-\,{a\over 2}\,)\,\cos\alpha\,\partial_{\theta}r
\,-\,V_{2\theta}\cos\alpha\,{\partial_{\varphi} r\over \sin\theta}
\,=0\,\,,\label{kappatres}\\\rc
&&V_{3\varphi}\,V_{2\theta}\,-\,
V_{3\theta}\,(\,V_{2\varphi}\,-\,{a\over 2}\,)\,-\,V_{1\varphi}\cos\alpha
\,\partial_{\theta}r\,+\,\rc\rc
&&\,\,\,\,\,\,\,\,\,\,\,\,\,\,\,\,\,\,\,\,\,\,\,
+\,\bigg(\,e^h\sin\alpha\,+\,(V_{1\theta}\,+\,{a\over 2}\,)\cos\alpha\bigg)\,
{\partial_{\varphi} r\over \sin\theta}\,=0\,\,,\label{kappacuatro}\\\rc
&&\sin\alpha \,V_{1\varphi}\partial_{\theta}r
\,-\,e^h\,V_{3\theta}
+\,\bigg(\,e^h\cos\alpha\,-\,(V_{1\theta}\,+\,{a\over 2}\,)\sin\alpha\,
\bigg)\,{\partial_{\varphi} r\over \sin\theta}
=0\,\,.
\label{kappacinco}
\eear
Moreover, if we want the kappa symmetry projection $\Gamma_{\kappa}\epsilon=\epsilon$ to 
coincide with the SUGRA projection, the ratio of the coefficients of the terms
with $\Gamma_{1}\,\hat\Gamma_{2}\,\epsilon$  and $\Gamma_{12}\,\epsilon$ must
be $\tan\alpha$, \ie\ one must have:

\beq
\tan\alpha\,=\,{c_{1\hat 2}\over c_{1 2}}\,\,.
\label{tanalpha}
\eeq
The explicit form of $c_{1 2}$ and $c_{1\hat 2}$ is:
\bear
c_{1 2}&=&
e^{2h}+V_{1\theta}\,V_{2\varphi}-
V_{2\theta}\,V_{1\varphi}-
{a\over 2}\,(\,V_{1\theta}\,-\,V_{2\varphi}\,)-{a^2\over 4}-
\cos\alpha V_{3\varphi}\,\partial_{\theta} r+
\cos\alpha V_{3\theta}{\partial_{\varphi}r\over\sin\theta}\,\,,\rc\rc
c_{1\hat 2}&=&
e^h\,(\,V_{2\varphi}\,-\,V_{1\theta}\,-a\,)\,-\,\sin\alpha
V_{3\varphi}\,\partial_{\theta} r\,+\,\sin\alpha
V_{3\theta}\,{\partial_{\varphi}r\over\sin\theta}\,\,.
\eear
Amazingly, except when $r$ is constant and takes values in the interval $0<r<\infty$ (see
section 4), eq. (\ref{tanalpha}) is a consequence of eqs. 
(\ref{kappauno})-(\ref{kappacinco}). Actually, by eliminating $V_{3\theta}$ of
eqs. (\ref{kappacuatro}) and (\ref{kappacinco}), and making use of eqs. 
(\ref{kappauno}) and (\ref{kappados}), one arrives at the following expression
of $\tan\alpha$:
\beq
\tan\alpha\,=\,{
e^h\,(\,V_{2\varphi}\,-\,V_{1\theta}\,-a\,)\over
e^{2h}\,+\,V_{1\theta}\,V_{2\varphi}\,-\,
V_{2\theta}\,V_{1\varphi}\,-\,
{a\over 2}\,(\,V_{1\theta}\,-\,V_{2\varphi}\,)\,-\,{a^2\over 4}}
\,\,.
\eeq
Notice that the terms of $c_{1\hat 2}$ ($c_{1 2}$) which do not contain
$\sin\alpha$ ($\cos\alpha$) are just the ones in the numerator (denominator) of
the right-hand side of this equation. It follows from this fact that eq. 
(\ref{tanalpha})  is satisfied if eqs. (\ref{kappauno})-(\ref{kappacinco})
hold. Moreover, by using the values of $\cos\alpha$ and $\sin\alpha$ given in
eq. (\ref{alpha}), one obtains the interesting relation:
\beq
\Big(\,1\,+\,a^2\,+\,4e^{2h}\,)\,(\,V_{1\theta}\,-\,V_{2\varphi}\,)=
4a\,\Big(\,V_{2\theta}^2\,+\,V_{1\theta}\,V_{2\varphi}\,-\,{1\over 4}
\,\Big)\,\,.
\label{condition}
\eeq
The system of eqs. (\ref{kappauno})-(\ref{kappacinco}) is rather involved and,
although it could seem at first sight very difficult and even hopeless to solve,
we will  be able to do it in some particular cases. Moreover, it
is interesting to notice that, by simple manipulations, one can obtain the
following expressions of the partial derivatives of $r$:
\bear
\partial_{\theta} r&=&-\cos\alpha V_{3\varphi}\,+\,\sin\alpha\,e^{-h}\,\,
\big[\,(V_{1\theta}\,+\,{a\over 2}\,)\,V_{3\varphi}\,-\,
V_{3\theta}\,V_{1\varphi}\,\big]\,\,,\rc\rc
\partial_{\varphi} r&=&\cos\alpha\sin\theta V_{3\theta}\,+\,
\sin\alpha\,\sin\theta e^{-h}\,\,
\big[\,(V_{2\varphi}\,-\,{a\over 2}\,)\,V_{3\theta}\,-\,
V_{3\varphi}\,V_{2\theta}\,\big]\,\,,
\label{rderivatives}
\eear
which will be very useful in our analysis.

\setcounter{equation}{0}
\section{Branes wrapped at fixed distance}
\medskip
In this section we will consider the possibility of wrapping the D5-branes at a fixed
distance $r>0$ from the origin. It is clear that, in this case, we have 
$\partial_{\theta}r\,=\,\partial_{\varphi}r=0$ and many of the terms on the left-hand
side of eqs. (\ref{kappauno})-(\ref{kappacinco}) cancel. 
Moreover $e^{h}$ is non-vanishing when $r>0$ and it can be
factored out in these equations. Thus, 
the equations (\ref{kappauno})-(\ref{kappacinco}) of kappa symmetry when the radial
coordinate $r$ is constant and non-zero reduce to:
\beq
V_{1\varphi}\,+\,V_{2\theta}\,=\,V_{3\varphi}\,=\,V_{3\theta}\,=\,0\,\,.
\eeq
From the equations $V_{3\varphi}\,=\,V_{3\theta}\,=\,0$ we obtain the following
differential equations for $\psi$
\beq
\partial_{\theta}\,\psi\,=\,-\cos\tilde\theta\,\partial_{\theta}\tilde\varphi\,\,,
\,\,\,\,\,\,\,\,\,\,\,\,\,\,\,\, 
\partial_{\varphi}\,\psi\,=\,-\cos\tilde\theta\,
\partial_{\varphi}\tilde\varphi\,-\,\cos\theta\,\,.
\label{eqpsi}
\eeq
The integrability condition for this system gives:
\beq
\partial_{\varphi}\tilde\theta\,
\partial_{\theta}\tilde\varphi\,-\,
\partial_{\theta}\tilde\theta\,
\partial_{\varphi}\tilde\varphi\,=\,
{\sin\theta\over\sin\tilde\theta}\,\,.
\label{integrability}
\eeq
By using this condition and the definition of the $V$'s (eq. (\ref{Vs})) one can prove
that
\beq
V_{1\theta}\,V_{2\varphi}\,-\,
V_{1\varphi}\,V_{2\theta}\,=\,-{1\over 4}\,\,.
\label{Vintegrability}
\eeq
Let us now define $\Delta$ as follows:
\beq
V_{2\varphi}\,-\,V_{1\theta}\,\equiv\,\,\Delta\,\,.
\label{Delta}
\eeq
By using the expression of the V's in terms of the angles, one can  combine eq.
(\ref{Delta})  and the condition 
$V_{1\varphi}\,+\,V_{2\theta}\,=\,0$ in the following matrix equation
\beq
\pmatrix{\cos\psi&\sin\psi\cr\cr
          -\sin\psi&\cos\psi}\,\,
\pmatrix{\sin\theta\partial_{\theta}\tilde\theta\,-\,
\sin\tilde\theta\partial_{\varphi}\tilde\varphi\cr\cr
\sin\theta\sin\tilde\theta\partial_{\theta}\tilde\varphi\,+\,
\partial_{\varphi}\tilde\theta}\,\,=\,\,
\pmatrix{-2\Delta\sin\theta\cr\cr 0}\,\,.
\eeq
Since the matrix appearing on the left-hand side is non-singular, we can multiply
by its inverse. By doing this one arrives at the following equations:
\bear
&&\partial_{\theta}\tilde\theta\,-\,{\sin\tilde\theta\over\sin\theta}\,
\partial_{\varphi}\tilde\varphi\,=\,-2\Delta\,\cos\psi\,\,,\rc\rc
&&{\partial_{\varphi}\tilde\theta\over \sin\theta}\,+\,
\sin\tilde\theta\partial_{\theta}\tilde\varphi\,=\,-2\Delta\,\sin\psi\,\,.
\label{nonCR}
\eear
Substituting the derivatives of $\tilde\theta$ obtained from the above equations into
the integrability condition (\ref{integrability}) we obtain after some calculation
\beq
\sin^2\tilde\theta\,
\Bigg(\,\partial_{\theta}\tilde\varphi\,+\,\Delta\,
{\sin\psi\over\sin\tilde\theta}\,\Bigg)^2\,+\,
{\sin^2\tilde\theta\over \sin^2\theta}\,\,
\Bigg(\,\partial_{\varphi}\tilde\varphi\,-\,\Delta\,
\cos\psi \,\,{\sin\theta\over\sin\tilde\theta}\,\Bigg)^2\,=\,
\Delta^2\,-\,1\,\,.
\label{squares}
\eeq
The right-hand side of eq. (\ref{squares}) is non-negative. Then, one obtains a bound for
$\Delta$:
\beq
\Delta^2\,\ge\,1\,\,.
\label{Deltabound}
\eeq
Notice that we have not imposed all the requirements of kappa symmetry. Indeed, it
still remains to check that the ratios between the coefficients $c_{12}$ and 
$c_{1\hat 2}$ is the one corresponding to the projection of the background. Using eq. 
(\ref{Vintegrability}) and the definition of $\Delta$ (eq. \ref{Delta})), one obtains:
\beq
c_{12}\,=\,e^{2h}\,+\,a\,{\Delta\over 2}\,-\,{a^2+1\over 4}\,\,,
\,\,\,\,\,\,\,\,\,\,\,\,\,\,\,\, \,\,\,\,\,\,\,\,\,\,\,\,\,\,\,\,
c_{1\hat 2}\,=\,e^h\,(\Delta-a)\,\,.
\eeq
Then, one must have:
\beq
\tan\alpha\,=\,{e^h\,(\Delta-a)\over 
e^{2h}\,+\,a\,{\Delta\over 2}\,-\,{a^2+1\over 4}}\,=\,
-{ae^{h}\over e^{2h}+{1-a^2\over 4}}\,\,,
\label{alphaDelta}
\eeq
where we have used the values of $\sin\alpha$ and $\cos\alpha$ given in the appendix A
(eq. A.17). If $e^h$ is nonzero (and finite), we can factor it out  in eq. 
(\ref{alphaDelta}) and obtain the following expression of $\Delta$:
\beq
\Delta= {2a\over 1+a^2+4e^{2h}}\,\,.
\eeq
Notice that $\Delta$ depends only on the coordinate $r$ and is a monotonically
decreasing function such that $0<\Delta<1$ for $0<r<\infty$ and
\beq
\lim_{r\rightarrow0}\,\,\Delta=1\,\,,
\,\,\,\,\,\,\,\,\,\,\,\,\,\,\,\, 
\lim_{r\rightarrow\infty}\,\,\Delta\,=\,0\,\,.
\eeq
As $\Delta<1$, the bound  (\ref{Deltabound}) is not satisfied and, thus, there is no
solution to our equations for $0<r<\infty$. Notice that this was to be expected from the
lack of moduli space of the ${\cal N}=1$ theories.

Let us now consider the possibility of placing the brane probe at $r\rightarrow\infty$.
Notice that in this case eq. (\ref{alphaDelta}) is satisfied for any finite value of
$\Delta$. However, the value $\Delta=1$ is special since, in this case, the 
right-hand side of eq. (\ref{squares}) vanishes and we obtain two 
equations that determine
the derivatives of $\varphi$, namely:
\beq
\partial_{\theta}\tilde\varphi\,=\,-{\sin\psi\over\sin\tilde\theta}\,\,,
\,\,\,\,\,\,\,\,\,\,\,\,\,\,\,\, \,\,\,\,\,\,\,\,\,\,\,\,\,\,\,\, 
\partial_{\varphi}\tilde\varphi\,=\,
\cos\psi \,\,{\sin\theta\over\sin\tilde\theta}
\label{partialvarphi}
\eeq
Using these equations into the system (\ref{nonCR}) for $\Delta=1$ one gets the
following equations for the derivatives of $\tilde\theta$:
\beq
\partial_{\theta}\tilde\theta\,=\,-\cos\psi\,\,,
\,\,\,\,\,\,\,\,\,\,\,\,\,\,\,\, \,\,\,\,\,\,\,\,\,\,\,\,\,\,\,\, 
\partial_{\varphi}\tilde\theta\,=\,-\sin\theta\sin\psi\,\,,
\label{partialtheta}
\eeq
and, similarly,  the equations  (\ref{eqpsi}) for $\psi$ become:
\beq
\partial_{\theta}\psi\,=\,\sin\psi\cot\tilde\theta\,\,,
\,\,\,\,\,\,\,\,\,\,\,\,\,\,\,\, \,\,\,\,\,\,\,\,\,\,\,\,\,\,\,\, 
\partial_{\varphi}\psi\,=\,-\sin\theta\cot\tilde\theta\,\cos\psi\,-\,
\cos\theta
\label{partialpsi}
\eeq
The equations (\ref{partialvarphi}) and (\ref{partialtheta}) can be regarded 
as coming from the following identifications of the frame forms in the 
$(\theta,\varphi)$ and $(\tilde\theta,\tilde\varphi)$ spheres:
\beq
\pmatrix{d\tilde\theta\cr\cr \sin\tilde\theta d\tilde\varphi} \,=\,
\pmatrix{\cos\psi&-\sin\psi\cr\cr\sin\psi&\cos\psi}\,
\pmatrix{-d\theta\cr\cr \sin\theta d\varphi}
\label{eqnsforms}
\eeq
The differential equations (\ref{partialpsi}) are just the integrability conditions of
the system (\ref{eqnsforms}). Another interesting observation is that one can prove by
using the differential eqs. (\ref{partialvarphi})-(\ref{partialpsi}) that the pullbacks
of the su(2) left-invariant one-forms are
\beq
P[w^1]\,=\,-d\theta\,\,,
\,\,\,\,\,\,\,\,\,\,\,\,\,\,\,\, \,\,\,\,\,\,\,\,\,\,\,\,\,\,\,\, 
P[w^2]\,=\,\sin\theta d\varphi\,\,,
\,\,\,\,\,\,\,\,\,\,\,\,\,\,\,\, \,\,\,\,\,\,\,\,\,\,\,\,\,\,\,\, 
P[w^3]\,=\,-\cos\theta d\varphi\,\,.
\label{pullbacksw}
\eeq

Let us try to find a solution of the differential equations
(\ref{partialvarphi})-(\ref{partialpsi}) in which 
$\tilde\theta=\tilde\theta(\theta)$ and 
$\tilde\varphi=\tilde\varphi(\varphi)$. The vanishing of
$\partial_{\varphi}\tilde\theta$ and $\partial_{\theta}\tilde\varphi$ immediately leads
to $\sin\psi=0$ or $\psi=0,\pi\,\,\,( \mod \,\,2\pi)$. Thus $\psi$ is constant in
this case. Let us put
$\cos\psi=\eta\,=\,\pm 1$. The vanishing of  $\partial_{\theta}\psi$ is automatic,
whereas the condition  $\partial_{\varphi}\psi\,=\,0$ leads to a relation between 
$\tilde\theta$ and $\theta$:
\beq
\cot\tilde\theta\,=\,-\eta\cot\theta
\eeq
In the case $\psi=0$, one has $\eta=1$ and the previous relation yields 
$\tilde\theta=\pi-\theta$. Notice that this relation is in agreement with
the first equation in eq. (\ref{partialtheta}). Moreover, the second equation in
(\ref{partialvarphi}) gives $\tilde\varphi=\varphi$. Similarly one can solve the
equations for  $\psi=\pi$. 
The solutions in these two cases are
just the ones used in ref. \cite{BertMer} in the calculation of the beta function  (with
some correction in the $\psi=0$ case to have the correct range of $\theta$ and
$\tilde\theta$), namely:
\bear
&&\tilde\theta\,=\,\pi-\theta\,\,,
\,\,\,\,\,\,\,\,\,\,\,\,\,
\tilde\varphi\,=\,\varphi\,\,,
\,\,\,\,\,\,\,\,\,\,\,\,\,
\psi=0 \,\,( \mod \,\,2\pi)\,\,,\rc\rc
&&\tilde\theta\,=\,\theta\,\,,
\,\,\,\,\,\,\,\,\,\,\,\,\,
\tilde\varphi\,=\,2\pi-\varphi\,\,,
\,\,\,\,\,\,\,\,\,\,\,\,\,
\psi=\pi \,\,( \mod \,\,2\pi)\,\,
\label{BMembeddings}
\eear

It follows from our results that the  embedding of ref. \cite{BertMer} is only
supersymmetric asymptotically when $r\rightarrow\infty$. In this sense, although it is
somehow distinguished,   it is not unique since for any embedding such that the $V$'s are
finite when
$r\rightarrow\infty$, the determinant of the induced metric diverges as $\sqrt{-g}\sim
e^{{3\phi\over 2}+2h}$ and the only term which survives in the equation
$\Gamma_{\kappa}\epsilon=\epsilon$ is the one with the matrix $\Gamma_{12}$, giving
rise to the same projection as the background for $r\rightarrow\infty$.

\setcounter{equation}{0}
\section{Worldvolume solitons (abelian case)}
\medskip
Let us consider the case $a=\alpha=0$ in the general equations of section 3.  From 
equations (\ref{kappauno})  and  (\ref{condition}) we get the following
(Cauchy-Riemann like) equations:
\beq
V_{1\theta}\,=\,V_{2\varphi}\,\,,
\,\,\,\,\,\,\,\,\,\,\,
V_{1\varphi}\,=\,-V_{2\theta}\,\,,
\label{VCR}
\eeq
whereas, from eq. (\ref{rderivatives}) we obtain that the derivatives of $r$
are given by:
\beq
r_{\theta}\,=\,-V_{3\varphi}\,\,,
\,\,\,\,\,\,\,\,\,\,\,\,\,\,\,\,\,\,\,\,\,\,\,
r_{\varphi}\,=\,\sin\theta\,V_{3\theta}\,\,,
\label{abrderiv}
\eeq
where $r_{\theta}\equiv \partial_{\theta} r$ and 
$r_{\varphi}\equiv \partial_{\varphi} r$. 
It can be easily demonstrated that, in this abelian case, the full set of
equations (\ref{kappauno})-(\ref{kappacinco})  collapses to the two pairs of
equations  (\ref{VCR}) and (\ref{abrderiv}). Notice that $c_{1\hat 2}=0$ when 
$a=\alpha=0$ and eq. (\ref{VCR}) holds and, thus, eq. (\ref{tanalpha}) is satisfied
identically. 

Let us study first the two equations (\ref{VCR}). By using the same technique as the one
employed in section 4 to derive eq. (\ref{nonCR}),  it can be shown easily that they
can be written as
\beq
\sin\theta\partial_{\theta}\tilde\theta\,=\,
\sin\tilde\theta\partial_{\varphi}\tilde\varphi\,\,,
\,\,\,\,\,\,\,\,\,\,\,\,\,\,\,\,\,\,\,\,\,\,
\partial_{\varphi}\tilde\theta\,=\,-
\sin\theta\sin\tilde\theta\partial_{\theta}\tilde\varphi\,\,.
\label{CR}
\eeq
In order to find the general solution of eq. (\ref{CR}), let us introduce a new set of
variables $u$ and $\tilde u$ as follows:
\beq
u\,=\,\log\tan{\theta\over 2}\,\,,
\,\,\,\,\,\,\,\,\,\,\,\,\,\,\,
\tilde u\,=\,\log\tan{\tilde\theta\over 2}\,\,.
\label{uvariables}
\eeq
Then,  eq. (\ref{CR}) can be written as the Cauchy-Riemann equations in the 
$(u,\varphi)$ and $(\tilde u,\tilde\varphi)$ variables, namely:
\beq
{\partial \tilde u\over \partial u}\,=\,
{\partial \tilde \varphi\over \partial \varphi}\,\,,
\,\,\,\,\,\,\,\,\,\,\,\,\,\,\,
{\partial \tilde u\over \partial \varphi}\,=\,-
{\partial \tilde \varphi\over \partial u}\,\,.
\label{CRu}
\eeq
Since $u, \tilde u\in (-\infty, +\infty)$ and 
 $\varphi, \tilde \varphi\in (0, 2\pi)$, the above equations are  the
Cauchy-Riemann equations in a band. The general solution of these equations is
of the form:
\beq
\tilde u+i\tilde\varphi\,=\,f(u+i\varphi)\,\,,
\eeq
where $f$ is an arbitrary function. Given any function $f$, it is clear that the above
equation provides the general solution $\tilde\theta (\theta, \varphi)$ and 
$\tilde\varphi (\theta, \varphi)$ of the system (\ref{CR}).

Let us turn now to the analysis of the system of equations (\ref{abrderiv}), which
determines the radial coordinate $r$. By using the explicit values of 
$V_{3\varphi}$ and $V_{3\theta}$, these equations can be written as:
\bear
r_{\theta}&=&-{1\over 2\sin\theta}\,\partial_{\varphi}\,\psi\,-\,
{1\over 2}\,{\cos\tilde\theta\over \sin\theta}\,\partial_{\varphi}\tilde\varphi
-\,{1\over 2}\cot\theta
\,\,,\rc\rc
r_{\varphi}&=&{\sin\theta\over 2}\,\partial_{\theta}\,\psi\,+\,
{\sin\theta\over 2}\,\cos\tilde\theta\,\partial_{\theta}\tilde\varphi\,\,,
\eear
where $\tilde\theta (\theta, \varphi)$ and $\tilde\varphi (\theta, \varphi)$ are
solutions of eq. (\ref{CR}). In terms of the derivatives
 with respect to variable $u$ defined
above ($\sin\theta\partial_{\theta}=\partial_u$), these equations become:
\bear
r_u&=&-{1\over 2}\,\,\partial_{\varphi}\psi\,-\,{1\over 2}\,\cos\tilde\theta\,
\partial_{\varphi}\tilde\varphi\,-\,{1\over 2}\,\cos\theta\,\,,\rc\rc
r_{\varphi}&=&{1\over 2}\,\,\partial_{u}\psi\,+\,{1\over 2}\,
\cos\tilde\theta\,\partial_{u}\tilde\varphi\,\,.
\label{eqr}
\eear
The integrability condition of these equations is just 
$\partial_{\varphi}r_u=\partial_{u}r_{\varphi}$. As any solution 
$(\tilde\theta, \tilde\varphi)$ of the  Cauchy-Riemann equations (\ref{CR}) satisfies:
\beq
\partial_{\varphi}\tilde\theta\partial_{\varphi}\tilde\varphi\,=\,-
\partial_{u}\tilde\theta\partial_{u}\tilde\varphi\,\,,
\eeq
and, since $\tilde\varphi$, being a solution of the Cauchy-Riemann equations, is
harmonic in $(u,\varphi)$, it follows that 
$\partial_{\varphi}r_u=\partial_{u}r_{\varphi}$ if and only if $\psi$ is
also harmonic in $(u,\varphi)$, \ie\ the differential equation for $\psi$ is just the
Laplace equation in the $(u,\varphi)$ plane, namely
\beq
\partial_{\varphi}^2\psi\,+\,\partial_u^2\psi\,=\,0\,\,.
\label{laplacepsi}
\eeq
Remarkably, the form of
$r(\theta,\varphi)$ can be obtained in general. Let us define:
\beq
\Lambda(\theta,\varphi)\,=\,\int_{0}^{\varphi} d\varphi \sin\theta
\partial_{\theta}\psi(\theta,\varphi)\,-\,
\int {d\theta\over \sin\theta}\,\, \partial_{\varphi}\psi(\theta,0)\,\,,
\eeq
It follows from this definition and the fact that $\psi$ is harmonic 
in $(u,\varphi)$ that $\psi$ and $\Lambda$ also satisfy the Cauchy-Riemann equations:
\beq
{\partial \Lambda\over \partial \varphi}\,=\,
{\partial \psi\over \partial u}\,\,,
\,\,\,\,\,\,\,\,\,\,\,\,\,\,\,\,\,\,\,\,\,
{\partial \Lambda\over \partial u}\,=\,-
{\partial \psi\over \partial \varphi}\,\,.
\label{CRpsi}
\eeq
Thus $\psi$ and $\Lambda$ are conjugate harmonic functions, \ie\ $\psi+i\Lambda$
is an analytic function of $u+i\varphi$. Notice that given $\Lambda$ one can
obtain $\psi$ by integrating the previous differential equations. It can be
checked by using the Cauchy-Riemann equations that the derivatives of $r$, as
given by the right hand side of eq. (\ref{eqr}), can be written as
$r_{\theta}=\partial_{\theta} F$, $r_{\varphi}=\partial_{\varphi} F$, where:
\beq
F(\theta,\varphi)\,=\,{1\over 2}\,\,\bigg[\,\Lambda(\theta,\varphi)\,-\,
\log\big(\sin\theta\sin\tilde\theta(\theta,\varphi)\big)\,\bigg]\,\,.
\eeq
Therefore, it follows that:
\beq
e^{2r}\,=\,C\,\,\,{e^{\Lambda(\theta,\varphi)}\over 
\sin\theta\sin\tilde\theta(\theta,\varphi)}\,\,,
\label{solution}
\eeq
with $C$ being a constant. We will make use of this amazingly simple expression 
to derive the equation of some particularly interesting embeddings.

\subsection{$n$-Winding solitons}

First of all, let us consider the particular class of solutions of the Cauchy-Riemann
eqs. (\ref{CRu}):
\beq
\tilde u+i\tilde\varphi\,=\,n(u+i\varphi)\,+\,{\rm constant}\,\,,
\eeq
where $n$ is an integer and the constant is complex. In terms of the original
variables:
\beq
\tan{\tilde\theta\over 2}\,=\,\tilde c\,
\Bigg(\,\tan{\theta\over 2}\,\Bigg)^{n}\,\,,
\,\,\,\,\,\,\,\,\,\,\,\,\,\,\,\,\,
\tilde\varphi\,=\,n\,\varphi\,+\,\varphi_0\,\,,
\label{VCRsol}
\eeq
with $\tilde c$ and $\varphi_0$ being constants. 
It is clear that in this solution the $\tilde\varphi$ coordinate of the probe wraps $n$
times the $[0,2\pi]$ interval as $\varphi$ varies between $0$ and $2\pi$. Let us now assume
that the coordinate $\psi$ is constant, \ie\ $\psi=\psi_0$. It is clear from its definition
that the function $\Lambda(\theta,\varphi)$ is zero in this case. Moreover, 
by using the identities
\beq
\sin x\,=\,{2\tan{x\over 2}\over
1+\tan^2{x\over 2}}\,\,,
\,\,\,\,\,\,\,\,\,\,\,\,\,\,\,\,\,\,\,\,\,
\tan{x\over 2}\,=\,\sqrt{{1-\cos x\over 1+\cos x}}\,\,,
\eeq
one can prove that:
\beq
\sin\tilde\theta\,=\,2\sqrt c\,\,\,{(\sin\theta\,)^{n}\over
(\,1\,+\,\cos\theta\,)^n\,+\,c\,(\,1\,-\,\cos\theta\,)^n}\,\,,
\label{tildetheta}
\eeq
where $c=\tilde c^2$. 
After plugging this result  in eq. (\ref{solution}), one obtains  the explicit form
of the function  $r(\theta)$, namely:
\beq
e^{2r}\,=\,{e^{2r_{*}}\over 1+c}\,\,
{(\,1\,+\,\cos\theta\,)^n\,+\,c\,(\,1\,-\,\cos\theta\,)^n\over
(\sin\theta\,)^{n+1}}\,\,,
\label{abspikes}
\eeq
where $r_{*}=r(\theta=\pi/2)$. We will call $n$-winding embedding to the brane
configuration corresponding to eqs. (\ref{VCRsol}) and (\ref{abspikes}) for a constant
value of the angle $\psi$.

\begin{figure}
\centerline{\hskip -.8in \epsffile{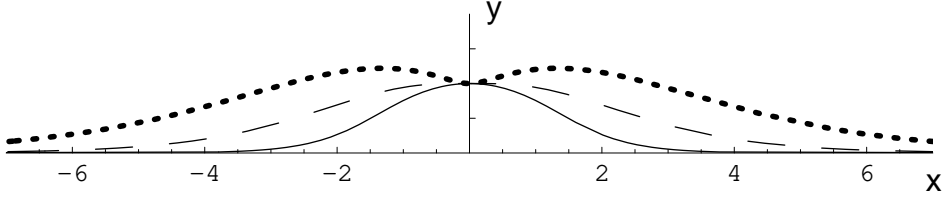}}
\caption{Curves $y=y(x)$ for three values of the winding number $n$: $n=0$ (solid line), 
$n=1$ (dashed curve) and $n=2$ (dotted line). These three curves correspond to $r_*=1$. }
\label{fig1}
\end{figure}

Let us pause for a moment to study the function (\ref{abspikes}). First of all it is
easy to verify that this function is invariant if we change $n\rightarrow -n$ and
$c\rightarrow 1/c$ (or equivalently changing $\theta\rightarrow \pi-\theta$ for the
same constant $c$). Actually, in what follows we shall take the integration constant
$c=1$ and thus we can restrict ourselves to the case in which $n$ is non-negative. 
In this $c=1$ case $r_*$ is the minimal separation between the brane probe and the
origin.
Another observation is that $r$ diverges for $\theta=0,\pi$,
which corresponds to the location of the spikes of the worldvolume solitons. Therefore
the supersymmetric embedding we have found is non-compact. Actually, it has the topology of
a cylinder whose compact direction is parametrized by $\varphi$. This cylinder connects
the two poles at $\theta=0,\pi$ of the $(\theta,\varphi)$ sphere at $r=\infty$ and passes
at a distance $r_*$ from the origin.

It is also interesting to discuss the symmetries of our
solutions. Recall that the angle $\psi$ is constant for our embeddings. Thus, it is clear
that one can shift it by an arbitrary constant $\epsilon$ as $\psi\to\psi+\epsilon$. This
U(1) symmetry corresponds to an isometry of the abelian background which
quantum-mechanically is broken to $\ZZ_{2N}$ as a consequence of the flux quantization of
the RR two-form potential
\cite{MN, KOW, GHP}. In the gauge theory side this isometry has been identified 
\cite{MN, KOW, GHP} with the $U(1)$ R-symmetry of the ${\cal N}=1$ SYM theory, which is
broken down to $\ZZ_{2N}$ by a field theory anomaly \cite{Affleck:1983mk}.
On the other hand, it is also clear that we have an additional U(1)
associated to constant shifts in
$\tilde\varphi$,  which are equivalent to a redefinition of $\varphi_0$ in eq.
(\ref{VCRsol}).

To visualize the shape of the
brane in these solutions it is rather convenient to introduce the following cartesian
coordinates $x$ and $y$
\beq
x\,=\,r\cos\theta\,\,,
\,\,\,\,\,\,\,\,\,\,\,\,\,\,\,
y\,=\,r\sin\theta\,\,.
\label{cartesian}
\eeq
In terms of $(x,y)$ the D5-brane embedding will be described by means of a curve
$y=y(x)$. Notice that $y\ge 0$, whereas $-\infty<x<+\infty$. The value of the coordinate
$y$ at $x=0$ is just $r_*$, \ie\ $y(x=0)=r_*$. Moreover, for large values of the
coordinate $x$, the function $y(x)\rightarrow 0$ exponentially as
\beq
y(x)\,\approx C\,|x|\,e^{-{2\over |n|+1}|x|}\,\,,
\,\,\,\,\,\,\,\,\,\,\,\,\,\,\,
(|x|\rightarrow \infty)\,\,,
\eeq
where $C$ is a constant. To illustrate this behaviour we have plotted in figure 1 the
curves $y(x)$ for three different values of the winding $n$ and the same value of 
$r_*$.

A particularly interesting
case is obtained when $n=\pm 1 $. By adjusting properly the constant $\varphi_0$ in
eq. (\ref{VCRsol}) the angular embedding reduces to:
\bear
&&\tilde\theta\,=\,\theta\,\,,
\,\,\,\,\,\,\,\,\,\,\,\,\,\,
\tilde\varphi\,=\,\varphi\,+\,{\rm constant}\,\,,
\,\,\,\,\,\,\,\,\,\,\,\,\,\,\,\,\,\,\,\,\,\,\,\,\,\,\,\,\,\,\,\,\,\,\,\,\,\,\,\,\,\,
(n=1)\,\,,\rc\rc
&&\tilde\theta\,=\,\pi-\theta\,\,,
\,\,\,\,\,\,\,\,\,\,\,\,\,\,
\tilde\varphi\,=\,2\pi-\varphi\,+\,{\rm constant}\,\,,
\,\,\,\,\,\,\,\,\,\,\,\,\,\,(n=-1)\,\,,
\label{BM}
\eear 
with $\psi$ being constant. 
These types of angular embeddings are similar to the ones  considered in ref.
\cite{BertMer} (although they are not the same, see eq. (\ref{BMembeddings}) ) 
and we will refer to them as
unit-winding embeddings. Notice that the two cases displayed in eq.  (\ref{BM}) represent
the two possible identifications of the
$(\theta,\varphi)$ and
$(\tilde\theta,\tilde\varphi)$ two-spheres. 

When $n=0$ the brane is wrapping the $(\theta,\varphi)$ sphere at constant values of
$\tilde\theta$ and $\tilde\varphi$, \ie\ one has:
\beq
\tilde\theta\,=\,{\rm constant}\,=\,\tilde\theta_0\,\,,
\,\,\,\,\,\,\,\,\,\,\,\,\,\,\,\,\,\,\,
\tilde\varphi\,=\,{\rm constant}\,=\,\tilde\varphi_0\,\,,
\,\,\,\,\,\,\,\,\,\,\,\,\,\,\,\,\,\,\,(n=0)
\eeq
We will refer to this case as zero-winding embedding \cite{DV}. 

One can verify that
the brane embeddings we have found are solutions of the probe equations of motion.
Actually, they are supersymmetric worldvolume solitons of the D5
brane probe. To illustrate this fact let us show that these configurations saturate a
BPS energy bound. To simplify matters, let us assume that the angular embedding is
the one displayed in eq. (\ref{VCRsol}) and let $r(\theta)$ be an arbitrary function. 
The Dirac-Born-Infeld (DBI) lagrangian density for a D5-brane with unit tension is:
\beq
{\cal L}\,=\,-e^{-\phi}\,\sqrt{-g_{st}}\,+\,P\,\Big[\,C_{(6)}\,\Big]\,\,,
\label{DBI}
\eeq
where $g_{st}$ is the determinant of the induced metric in the string frame
($g_{st}=e^{3\phi}\,g$) and  $P\,\Big[\,C_{(6)}\,\Big]$ is the pullback on the
worldvolume of the RR six-form of the background. The elements of the induced metric for
the $n$-winding solution along the angular coordinates are:
\bear
g_{\theta\theta}&=&e^{{\phi\over 2}}\,
\bigg(\,e^{2h}\,+\,{n^2\over 4}\,
{\sin^2\tilde\theta\over \sin^2\theta}\,+\,r_{\theta}^2\,\bigg)\,\,,\cr\cr
g_{\varphi\varphi}&=&e^{{\phi\over 2}}\,
\bigg(\,e^{2h}\,+\,{n^2\over 4}\,
{\sin^2\tilde\theta\over \sin^2\theta}\,+\,V_{3\varphi}^2\,\bigg)\sin^2\theta
\,\,.
\eear
From this expression one immediately obtains the determinant of the induced metric,
namely:

\beq 
\sqrt{-g}\,=\,e^{{3\phi\over 2}}\,\sin\theta\,\,
\sqrt{\bigg(\,e^{2h}\,+\,{n^2\over 4}\,
{\sin^2\tilde\theta\over \sin^2\theta}\,+\,V_{3\varphi}^2\,\bigg)\,
\bigg(\,e^{2h}\,+\,{n^2\over 4}\,
{\sin^2\tilde\theta\over \sin^2\theta}\,+\,r_{\theta}^2\,\bigg)}\,\,.
\eeq
Moreover, the pullback on the worldvolume of the two-form ${\cal C}$ is
\footnote{It is worth mentioning that the pullback of the RR two-form to the
worldvolume is
$$
P[C_{(2)}]\,=\,{\psi\over 4}\,d\varphi\wedge 
\big(\,n\sin\tilde\theta d\tilde\theta\,-\,\sin\theta d\theta\,\big)\,\,,
$$
where $\tilde\theta(\theta)$ is the function displayed in eq. (\ref{tildetheta}) . From
this expression it is straightforward to verify that the RR two-form flux through the
two-submanifold where we are wrapping our brane is
$$
\int\,P[C_{(2)}]\,=\,\pi\psi\,(|n|-1)\,\,,
$$
and thus it vanishes  iff $n=\pm 1$.
 }:
\beq
P\,\Big[\,{\cal C}\,\Big]\,=\,{e^{2\phi}\over 8}\,
(\,16e^{2h}\,\cos\theta\,-\,ne^{-2h}\,\cos\tilde\theta\,)\,r_{\theta}\,
d\varphi\wedge d\theta
\eeq
The hamiltonian density ${\cal H}$ for a static configuration is just 
${\cal H}\,=\,-{\cal L}$ or:
\bear
&&{\cal H}\,=\,e^{2\phi}\,\Bigg[\,\sin\theta\,
\sqrt{\bigg(\,e^{2h}\,+\,{n^2\over 4}\,
{\sin^2\tilde\theta\over \sin^2\theta}\,+\,V_{3\varphi}^2\,\bigg)\,
\bigg(\,e^{2h}\,+\,{n^2\over 4}\,
{\sin^2\tilde\theta\over \sin^2\theta}\,+\,r_{\theta}^2\,\bigg)}\,-\,\rc\rc\rc
&&\,\,\,\,\,\,\,\,\,\,\,\,\,\,\,\,\,\,\,
\,-\,{1\over 8}\,
(\,16e^{2h}\,\cos\theta\,-\,ne^{-2h}\,\cos\tilde\theta\,)\,r_{\theta}\,
\Bigg]\,\,,
\eear
It can be checked that, for an arbitrary function $r(\theta)$, one can write
${\cal H}$ as:
\beq
{\cal H}\,=\,{\cal Z}\,+\,{\cal S}\,\,,
\eeq
where ${\cal Z}$ is a total derivative:
\beq
{\cal Z}\,=\,-\partial_{\theta}\,\Bigg[\,
e^{2\phi}\,\bigg(e^{2h}\,\cos\theta\,+\,{n\over 4}\, \cos\tilde\theta\,
\bigg)\,\Bigg]\,\,,
\eeq
and ${\cal S}$ is non-negative:
\beq
{\cal S}\ge 0\,\,,
\eeq
with ${\cal S}= 0$ precisely when the BPS equations for the embedding are
satisfied. The expression of ${\cal S}$ is:
\bear
{\cal S}&=&\sin\theta\,e^{2\phi}\,\Bigg[\,
\sqrt{\bigg(\,e^{2h}\,+\,{n^2\over 4}\,
{\sin^2\tilde\theta\over \sin^2\theta}\,+\,V_{3\varphi}^2\,\bigg)\,
\bigg(\,e^{2h}\,+\,{n^2\over 4}\,
{\sin^2\tilde\theta\over \sin^2\theta}\,+\,r_{\theta}^2\,\bigg)}\,\,-\rc\rc
&&\,\,\,\,\,\,\,\,\,\,\,\,\,\,\,\,\,\,\,\,\,\,\,\,\,\,\,\,
-\bigg(\,e^{2h}\,+\,{n^2\over 4}\,
{\sin^2\tilde\theta\over \sin^2\theta}\,-\,V_{3\varphi}\,r_{\theta}\,\bigg)\,
\Bigg]\,\,.
\label{calS}
\eear
The BPS equation for $r$ in this case is $r_{\theta}=-V_{3\varphi}$ (see eq.
(\ref{abrderiv})). If this equation is satisfied,  the first term on the right-hand side
of eq.  (\ref{calS}) is a square root of a perfect square which cancels against the second
term of this equation. Moreover, it is easy to check that
the condition ${\cal S}\ge 0$ is equivalent to:
\beq
\big(r_{\theta}\,+\,V_{3\varphi}\,\big)^2\ge 0\,\,,
\eeq
which is obviously satisfied and reduces to an equality if and only if the BPS
equation for the embedding is satisfied. 

\subsection{(n,m)-Winding solitons}

The solutions found in the previous section are easily generalized if we allow the
angle $\psi$ to wind a certain number of times as the coordinate $\varphi$ varies from 
$\varphi=0$ to $\varphi=2\pi$. Recalling that $\psi$ ranges from $0$ to $4\pi$, let us
write the following ansatz for $\psi(\varphi)$:
\beq
\psi\,=\,\psi_0\,+\,2m\varphi\,\,,
\eeq
where $m$ is an integer. It is obvious that the above function satisfies the Laplace
equation (\ref{laplacepsi}). Moreover, its harmonic conjugate $\Lambda$ is immediately
obtained by solving the Cauchy-Riemann differential equations (\ref{CRpsi}), namely:
\beq
\Lambda\,=\,-2mu\,\,.
\eeq
In terms of the angle $\theta$, the above equation becomes:
\beq
e^{\Lambda}\,=\,{1\over \Big(\,\tan{\theta\over 2}\,\Big)^{2m}}\,\,.
\eeq
By plugging this result in eq. (\ref{solution}), and using the value of
$\sin\tilde\theta$ given in eq.  (\ref{tildetheta}), it is straightforward to obtain
the function $r(\theta)$ of the embedding. One gets:
\beq
e^{2r}\,=\,{e^{2r_{*}}\over 1+c}\,\,
{(\,1\,+\,\cos\theta\,)^n\,+\,c\,(\,1\,-\,\cos\theta\,)^n\over
\big[\tan{\theta\over 2}\big]^{2m}
(\sin\theta\,)^{n+1}}\,\,,
\label{nmsol}
\eeq
where, as in the $n$-winding case,   $r_{*}=r(\theta=\pi/2)$.

An interesting observation concerning the solution we have just found is that, by
choosing appropriately the winding number $m$, one of the spikes of the $m=0$
solutions at $\theta=0$ or $\theta=\pi$ disappears. Indeed if, for example,  $n$ is
nonnegative and we take $2m=n+1$, the function $r(\theta)$ is regular at $\theta=0$.
Similarly, also when $n\ge 0$, one can eliminate the spike at $\theta=\pi$ by choosing
$2m=-n-1$.

\subsection{Spiral solitons}

By considering more general solutions of the Cauchy-Riemann equations (\ref{CRu}) and 
(\ref{CRpsi}) we can obtain many more classes of supersymmetric configurations of the
brane probe. One of the questions one can address is whether or not one can have
embeddings in which $r$ is finite for all values of the angles. We will now see that
the answer to this question is yes, although the corresponding embeddings seem not to be
very interesting. To illustrate this point, let us see how we can find functions
$\psi$ and $\Lambda$ such that they make the radial coordinate of the $n$-winding
embedding finite at $\theta=0, \pi$. First of all, notice that, in terms of the
Cauchy-Riemann variables $u$ and $\tilde u$ defined in eq. (\ref{uvariables}), we have to
explore the behaviour of the embedding at $u, \tilde u\rightarrow \pm\infty$. Since
\beq
\sin\theta\,=\,{2e^{u}\over 1+e^{2u}}\,\,,
\,\,\,\,\,\,\,\,\,\,\,\,\,\,\,\,\,\,\,\,\,
\sin\tilde\theta\,=\,{2e^{\tilde u}\over 1+e^{2\tilde u}}\,\,,
\eeq
one has that $\sin\theta\rightarrow e^{-|u|}$,  $\sin\tilde\theta\rightarrow
e^{-|\tilde u|}$ as $u, \tilde u\rightarrow \pm\infty$. Then, the factors multiplying 
$e^{\Lambda}$ in eq. (\ref{solution}) diverge as $e^{|u|+|\tilde u|}$. In the
$n$-winding solution $|\tilde u|=|n|| u|$ and, therefore this divergence is of the type
$e^{(|n|+1)|u|}$. We can cancel this divergence by adding a
$\Lambda$ such that $e^{\Lambda}\rightarrow 0$ as $u\rightarrow\pm\infty$ in  such a
way that, for example,   $\Lambda+(|n|+1)|u|\rightarrow -\infty$. This is clearly
achieved by taking a function $\Lambda$ such that $\Lambda\rightarrow-u^2$. It is
straightforward to find an
analytic function in the $(u,\varphi)$ plane such that its imaginary part behaves as
$-u^2$ for $u\rightarrow\pm\infty$.  One can take
\beq
\psi+i\Lambda\,=\,-i(u+i\varphi)^2\,=\,2u\varphi\,-\,i(u^2-\varphi^2)\,\,.
\eeq
From this equation we can read the functions $\psi$ and $\Lambda$. In terms of 
$\theta$ and $\varphi$ they are:
\beq
\psi\,=\,2u\varphi\,=\,2\varphi\log\tan{\theta\over 2}\,\,,
\,\,\,\,\,\,\,\,\,\,\,\,\,\,\,\,\,\,\,\,\,
\Lambda\,=\,-u^2+\varphi^2\,=\,-\big(\,\log\tan{\theta\over 2}\,\big)^2\,+\,
\varphi^2\,\,.
\eeq
In this case $r\rightarrow 0$, $\psi\rightarrow\pm\infty$ as 
$\theta\rightarrow 0,\pi$, which means that we describe an infinite spiral which winds
infinitely in the $\psi$ direction. Notice that, although $r$ is always finite, the
volume of the two-submanifold is infinite due to this infinite winding. One can try other
alternatives to make the radial coordinate finite. In all the ones we have analyzed
one obtains the infinite spiral behaviour described above.

\setcounter{equation}{0}
\section{Worldvolume solitons (nonabelian case)}
\medskip
Let us consider the full nonabelian background and let us try to obtain solutions to
the kappa symmetry equations (\ref{kappauno})-(\ref{kappacinco}). Actually we will
restrict ourselves to the situations in which $r$ only depends on the angle $\theta$.
It can be easily checked that, in this case, only four of the five equations 
(\ref{kappauno})-(\ref{kappacinco}) are independent. As an independent set of equations
we will choose eqs. (\ref{kappauno}), (\ref{condition}) and
\bear
&&\partial_{\theta}r=-{e^{h}\,V_{3\varphi}\over
e^{h}\,\cos\alpha\,+\,(\,V_{2\varphi}\,-\,{a \over 2}\,)\,\sin\alpha}\,\,,
\label{nabuno}\\\rc
&&\sin\alpha V_{1\varphi}\,\partial_{\theta}r\,-\,e^{h}\,V_{3\theta}=0\,\,,
\label{nabdos}
\eear
which can be obtained from eqs. (\ref{kappados}) and (\ref{kappacinco}) after taking
$\partial_{\varphi} r=0$.

We will now try to find the non-abelian version of the solutions found in the
abelian theory for arbitrary winding $n$. With this purpose, let us consider the
following ansatz for $\tilde\varphi$:
\beq
\tilde\varphi(\theta,\varphi)\,=\,n\varphi\,+\,f(\theta)\,\,,
\eeq
while we shall assume that $\tilde\theta$, $\psi$ and $r$ are functions of
$\theta$ only. We will require that, in the asymptotic UV,
$\tilde\varphi\rightarrow n\varphi$. It this clear that in this ansatz
$\partial_{\varphi}\tilde\varphi\,=\,n$ and that 
$\partial_{\theta}\tilde\varphi\,=\,\partial_{\theta}f$. Moreover, from 
eq. (\ref{kappauno}) we can obtain the relation between
$\partial_{\theta}\tilde\varphi$ and 
$\partial_{\theta}\tilde\theta$, namely:
\beq
\partial_{\theta}\tilde\varphi\,=\,\tan\psi\,\,\bigg[\,
{\partial_{\theta}\tilde\theta\over \sin\tilde\theta}\,-\,
{n\over \sin\theta}\,\bigg]\,\,.
\label{rel}
\eeq
Using this value of $\partial_{\theta}\tilde\varphi$, we get the following
values of the $V$ functions:
\bear
&&V_{1\theta}\,=\,{1\over 2}\,\bigg[\,
{\partial_{\theta}\tilde\theta\over \cos\psi}\,-\,n\,
{\sin\tilde\theta\over \sin\theta}\,\,{\sin^2\psi\over
\cos\psi}\,\bigg]\,\,,\rc\rc
&&V_{1\varphi}\,=\,{n\over 2}\,{\sin\tilde\theta\over \sin\theta}\,\sin\psi
\,=\,-V_{2\theta}\,\,,\rc\rc
&&V_{2\varphi}\,=\,{n\over 2}\,{\sin\tilde\theta\over \sin\theta}\,\cos\psi
\,\,,\rc\rc
&&V_{3\theta}\,=\,{1\over 2}\,\partial_{\theta}\psi\,+\,
{1\over 2}\,\cot\tilde\theta\,\tan\psi\,\partial_{\theta}\tilde\theta\,-\,
{n\over 2}\,\tan\psi\,{\cos\tilde\theta\over\sin\theta}\,\,,\rc\rc
&&V_{3\varphi}\,=\,{n\over 2}\,{\cos\tilde\theta\over \sin\theta}\,+\,
{1\over 2}\,\cot\theta\,\,.
\label{nabVs}
\eear
By using these values in eq. (\ref{condition}), one gets the value of 
$\partial_{\theta}\tilde\theta$ in terms of the other variables:
\beq
\partial_{\theta}\tilde\theta\,=\,
{n\sin\tilde\theta\cosh 2r-\sin\theta\cos\psi\over 
\sin\theta\cosh 2r - n\sin\tilde\theta\cos\psi}\,\,.
\label{eqtildetheta}
\eeq
On the other hand, by combining the two equations 
(\ref{rel}) and (\ref{eqtildetheta}), we obtain:
\beq
\partial_{\theta}\,\tilde\varphi\,=\,
{n^2\sin^2\tilde\theta\,-\,\sin^2\theta\over
\sin\theta\sin\tilde\theta \,
(\,\sin\theta\cosh 2r\,-\,n\sin\tilde\theta\cos\psi\,)}\,\sin\psi\,\,.
\label{eqtildevarphi}
\eeq
Moreover, plugging the values of
$V_{1\varphi}$, $V_{2\varphi}$, $V_{3\theta}$ and $V_{3\varphi}$ in eqs. 
(\ref{nabuno}) and (\ref{nabdos}) we can obtain the values of the
derivatives of $\psi$ and $r$. The result is:
\bear
\partial_{\theta}\psi&=&
{n\cot\theta\sin\tilde\theta+\cot\tilde\theta\sin\theta\over
\sin\theta\cosh2r-n\sin\tilde\theta\cos\psi}\,\,\sin\psi\,\,,\rc\rc
\partial_{\theta}r&=&-{1\over 2}\,\,
{n\cos\tilde\theta+\cos\theta\over 
\sin\theta\cosh2r- n\sin\tilde\theta\cos\psi}\,\,\sinh 2r\,\,.
\label{eqpsiyr}
\eear

It follows from eq. (\ref{eqtildetheta}) that, asymptotically in the UV, 
$\sin\theta\,\partial_{\theta}\tilde\theta\rightarrow n\sin\tilde\theta$. In
order to fulfill this relation for arbitrary $r$ it is easy to see from 
eq. (\ref{eqtildetheta}) that one must have $(n\sin\tilde\theta)^2=\sin^2\theta$, which
only happens for $n=\pm 1$ and $\sin\theta=\sin\tilde\theta$. Noticing that for 
these values one has $\partial_{\theta}\tilde\theta\,=\,\pm 1$, one is finally
led to the two  possibilities of eq. (\ref{BM}): $\tilde\theta=\theta$ for $n=1$ and
$\tilde\theta=\pi-\theta$ for $n=-1$. Notice that in the two cases of eq.  (\ref{BM}) this
equation implies that 
$\partial_{\theta}\,\tilde\varphi=0$ and thus when $n=\pm 1$ the angular identifications of
the abelian unit-winding embeddings (eq. (\ref{BM})) also solve the non-abelian equations
(\ref{eqtildetheta}) and  (\ref{eqtildevarphi}) for
all $r$.

For a general value of $n$ one has  that asymptotically in the UV
$\partial_{\theta}\tilde\varphi\rightarrow 0$ and 
$\partial_{\theta}\psi\rightarrow 0$, as in the abelian solutions. Moreover 
it follows from eqs. (\ref{eqtildevarphi}) and (\ref{eqpsiyr}) that
$\tilde\varphi$ and $\psi$ can be kept constant for all $r$ if $\sin\psi=0$,
\ie\ when $\psi=0,\pi\,\,\,\mod\,\,2\pi$. For this values of $\psi$ the equations
simplify and, although we will not attempt to do it here, 
one could try to integrate numerically the equations of
$\tilde\theta$ and $r$. It is however interesting to point out that, contrary to what
happens in the abelian $n$-winding solution, the angle $\psi$ cannot be an arbitrary
constant for the non-abelian probes. As we will argue below, this is a geometrical
realization of the  breaking of the R-symmetry of the corresponding ${\cal N}=1$ SYM
theory in the IR. On the contrary, the angle
$\tilde\varphi$ can take an arbitrary constant value, as in the abelian solution.

\subsection{Non abelian unit-winding solutions}

Let us now obtain the non-abelian generalization of the unit-winding solutions. First of
all we define
\beq
\eta\,=\,n=\pm 1\,\,.
\eeq
We have already noticed that for unit-winding embeddings the values of $\tilde\theta$
and $\tilde\varphi$ displayed in eq. (\ref{BM}) solve the non-abelian differential
equations (\ref{eqtildetheta}) and (\ref{eqtildevarphi}). Therefore,  
let us try to find a solution in the nonabelian theory in which the 
embedding of the $(\tilde\theta, \tilde\varphi)$ coordinates
is the same as in the abelian theory, \ie\ as in eq. (\ref{BM}).  For this type
of embeddings  $\sin\tilde\theta=\sin\theta$, $\partial_{\theta}\tilde\theta\,=\,\eta$ and
eq. (\ref{nabVs}) reduces to:
\bear
&&V_{1\theta}\,=\,{\eta\cos\psi\over 2}\,\,,
\,\,\,\,\,\,\,\,\,\,\,\,\,\,\,\,\,\,\,\,\,\,\,\,\,\,
V_{1\varphi}\,=\,{\eta\sin\psi\over 2}\,\,,\rc\rc
&&V_{2\theta}\,=\,-{\eta\sin\psi\over 2}\,\,,
\,\,\,\,\,\,\,\,\,\,\,\,\,\,\,\,\,\,\,\,\,\,
V_{2\varphi}\,=\,{\eta\cos\psi\over 2}\,\,,\rc\rc
&&V_{3\theta}\,=\,{\psi_{\theta}\over 2}\,\,,
\,\,\,\,\,\,\,\,\,\,\,\,\,\,\,\,\,\,\,\,\,\,\,\,\,\,\,\,\,\,\,\,\,\,\,\,\,\,
V_{3\varphi}\,=\,\cot\theta\,\,,
\eear
where we have denoted $\psi_\theta\equiv\partial_{\theta}\,\psi$. As a check, notice that
$V_{1\theta}$,  $V_{1\varphi}$, $V_{2\theta}$ and $V_{2\varphi}$ satisfy eqs. (\ref{VCR}).
It follows from eq.  (\ref{condition}) that they must also satisfy
\beq
V_{1\theta}^2\,+\,V_{1\varphi}^2\,=\,{1\over 4}\,\,,
\,\,\,\,\,\,\,\,\,\,\,\,\,\,\,\,\,\,\,\,\,\,
V_{2\theta}^2\,+\,V_{2\varphi}^2\,=\,{1\over 4}\,\,,
\eeq
which indeed they verify. Moreover, by substituting $\sin\tilde\theta=\sin\theta$ and
$\cos\tilde\theta=\eta\cos\theta$ in eq. (\ref{eqpsiyr}), we obtain the following
differential equations for $\psi(\theta)$ and $r(\theta)$:
\bear
\psi_\theta&=&-{2\eta\sin\psi\over \sinh 2r}\,\,r_\theta\,\,,\rc\rc
r_\theta&=&-{\cot\theta\over 
\cosh 2r\,-\,\eta\cos\psi}\,\,
\sinh 2r\,\,.
\label{nabBPS}
\eear
These equations can be integrated with the result:
\bear
&&\Bigg(\,\tan\,{\psi\over 2}\,\Bigg)^{\eta}\,=\,A\,\coth r\,\,,\rc\rc
&& {\sinh r\over \sqrt{A^2+\tanh^2r}}\,=\,{C\over \sin\theta}\,\,, 
\label{BMnoab}
\eear
where $A$ and $C$ are constants of integration. Eq. (\ref{BMnoab}), together with eq.
(\ref{BM}), determines the unit-winding embeddings of the probe in the non-abelian
background. 
Notice that, as in the corresponding
abelian solution, $r$ diverges when $\theta=0,\pi$, \ie\ the brane probe extends
infinitely in the radial direction.  On the other hand,
it is also instructive to explore the $r\rightarrow\infty$ limit of the
solution (\ref{BMnoab}). First of all, it is clear that when $r\rightarrow\infty$ the angle
$\psi$ reaches asymptotically  a constant value $\psi_0$, given by
\begin{figure}
\centerline{\hskip -.8in \epsffile{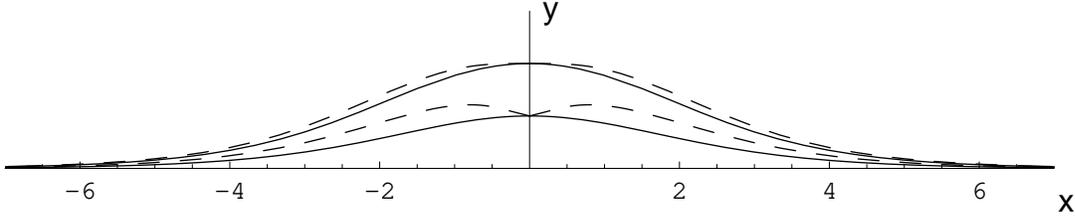}}
\caption{Comparison between the non-abelian (solid line) and abelian (dashed line)
unit-winding embeddings for the same value of $r_*$ . The non-abelian embedding is the
one corresponding to eq. (\ref{noabflavor}) and the abelian one is that given in 
eq. (\ref{abspikes}) 
for $n=1$ and $c=1$.  The curves for two different values of $r_*$
($r_*=0.5$ and $r_*=1$) are shown. The variables $(x,y)$ are the ones defined in eq.
(\ref{cartesian}). }
\label{fig2}
\end{figure}

\beq
\cos\psi_0\,=\,{1-A^2\over 1+A^2}\,\,\eta\,\,.
\eeq
Moreover, when $r\rightarrow\infty$ the function $r(\theta)$ displayed in eq.
(\ref{BMnoab}) becomes, after a proper identification of  the integration constants,
exactly the one written in eq. (\ref{abspikes}) for $n=\pm 1$ and $c=1$.
Notice that  the angle $\psi$ in the  embedding (\ref{BMnoab})
is not constant in general. Actually, only when $A=0$ or $A=\infty$ the coordinate
$\psi$ remains constant and equal to $0,\pi\,\,\mod\,2\pi$ ($\cos\psi=\eta$ for $A=0$ and
$\cos\psi=-\eta$ in the case
$A=\infty$). It is interesting to write the dependence of $r$ on $\theta$ in these two
particular cases. When $A=\infty$ the solution is

\bear
\tilde\theta&=&\cases{\theta, &if $\,\,\eta=+1\,,$\cr
                        \pi-\theta, &if $\,\,\eta=-1\,,$ }\,\,,
\,\,\,\,\,\,\,\,\,\,\,\,
\tilde\varphi\,=\,\cases{\varphi+{\rm constant}, &if $\,\,\eta=+1\,,$\cr
                        2\pi-\varphi+{\rm constant},  &if $\,\,\eta=-1$ }\,\,,\rc\rc
\psi&=&\cases{\pi,3\pi, &if $\,\,\eta=+1\,,$\cr
               0,2\pi, &if $\,\,\eta=-1\,,$ }\,\,,
\,\,\,\,\,\,\,\,\,\,\,\,
\sinh r\,=\,{\sinh r_*\over \sin\theta}\,\,\,,
\label{noabflavor}
\eear
where $r_*$ is the minimal value of $r$ (\ie\ $r_*=r(\theta=\pi/2)$) and we have also
displayed the angular part of the embedding. Notice that, for a given sign of the
winding number $\eta$, only two  values of $\psi$ are possible. Thus, in this
solution, the U(1) symmetry of shifts in $\psi$ is broken to a $\ZZ_2$ symmetry. This
will be interpreted in section 7 as the realization, at the level of the brane probe, of
the R-symmetry breaking of the gauge theory.

To have a better understanding of the solution (\ref{noabflavor}) we have plotted it
in figure 2
in terms of the variables $(x,y)$ defined in eq. (\ref{cartesian}). For comparison we
have also plotted the abelian solution corresponding to the same value of 
$r_*$. In this figure the embeddings for two different values of the minimal
radial distance $r_*$ are shown. When $r_*$ is large enough ($r_*\ge 2$) the two curves
become practically identical. 

Let us now have a look at the case of the $A=0$  embeddings. The function $r(\theta)$ in
this case can be read from eq. (\ref{BMnoab}), namely:
\beq
\cosh r\,=\,{C\over \sin\theta}\,\,\,.
\label{nabcosh}
\eeq
We have plotted in figure 3 the profile for these embeddings in terms of the variables
$(x,y)$ of eq. (\ref{cartesian}). Notice that, when C is in the interval $(1,\infty)$ it
can be parametrized as $C=\cosh r_*$, with $r_*>0$ being the minimal radial distance
between the probe and the origin. On the contrary, when $C$ lies in the interval $[0,1]$
the brane reaches the origin when $\sin\theta=C$. We have thus, in this case, a
one-parameter family of configurations which pass through the origin.

\begin{figure}
\centerline{\hskip -.8in \epsffile{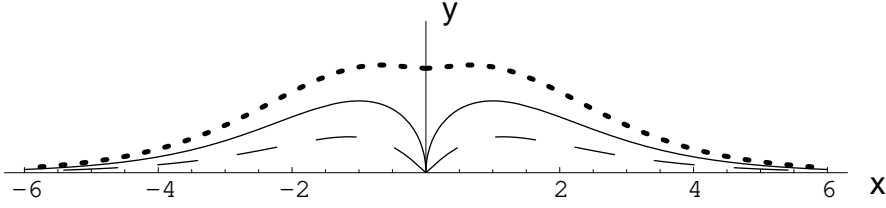}}
\caption{ Graphic representation of the unit winding embedding 
of eq. (\ref{nabcosh}) for three values of
the constant $C$: $C=0.5$ (dashed line), $C=1$ (solid line) and $C=1.5$ (dotted line).
The variables $(x,y)$ are the ones defined in eq.
(\ref{cartesian}). }
\label{fig3}
\end{figure}

As in their abelian counterparts,
these worldvolume solitons for the non-abelian background saturate an energy bound. In
order to prove this fact,  let us define
\beq
D\,\equiv\,\coth 2r\,-\,\eta\,{\cos\psi\over \sinh 2r}\,\,.
\eeq
Notice that $D\ge 0$ for any real $\psi$ and $r$. Moreover the equation for $r(\theta)$
can be written as
$r_{\theta}=-\cot\theta/D$. For arbitrary functions $r(\theta)$ and 
$\psi(\theta)$ the hamiltonian density takes the form:
\bear
{\cal H}\,=\,e^{2\phi}\sin\theta &\Bigg[&
\sqrt{\Big(r-r_{\theta}\cot\theta\Big)^2\,+\,
rD\,\Big(r_{\theta}+{\cot\theta\over D}\,\Big)^2\,+\,
{rD\over 4}\Big(\psi_{\theta}\,-\,{2\eta\sin\psi\over D\sinh 2r}\,\cot\theta
\,\Big)^2}\,-\rc\rc
&&-\,\Big(\,2e^{2h}\,-\,{1\over 8}\,(a^2-1)^2\,e^{-2h}\,
\,\Big)\,\cot\theta \,\,r_{\theta}\Bigg]\,\,.
\eear

It can be verified that ${\cal H}$ can be written as 
${\cal H}={\cal Z}+{\cal S}$, where
\beq
{\cal Z}=-{d\over d\theta}\,\Big[\,e^{2\phi}\,r\cos\theta\,\Big]\,\,.
\eeq
(this is the same value as in the abelian soliton for $n=1$). The expression of 
${\cal S}$ is:
\bear
{\cal S}\,=\,e^{2\phi}\sin\theta &\Bigg[&
\sqrt{\Big(r-r_{\theta}\cot\theta\Big)^2\,+\,
rD\,\Big(r_{\theta}+{\cot\theta\over D}\,\Big)^2\,+\,
{rD\over 4}\Big(\psi_{\theta}\,-\,{2\eta\sin\psi\over D\sinh 2r}\,\cot\theta
\,\Big)^2}\,-\rc\rc
&&\,-\Big(\,r\,-\,r_{\theta}\cot\theta\,\big)\,\Bigg]\,\,.
\label{nabcalS}
\eear
As in the abelian case, if the BPS equations (\ref{nabBPS}) are satisfied, the square
root on eq. (\ref{nabcalS}) can be exactly evaluated and ${\cal S}$ vanishes. Furthermore,
one can easily check that ${\cal S}\ge 0$ is equivalent to the condition
\beq
rD\,\Big(r_{\theta}+{\cot\theta\over D}\,\Big)^2\,+\,
{rD\over 4}\Big(\psi_{\theta}\,-\,{2\eta\sin\psi\over D\sinh 2r}\,\cot\theta
\,\Big)^2\,\ge\,0\,\,,
\eeq
which, since $D\ge 0$, is trivially satisfied for any functions $r(\theta)$ and
$\psi(\theta)$. Moreover, as $r-r_{\theta}\cot \theta\ge 0$ for the solution
of the BPS equations, it follows that our BPS embedding saturates the bound.

\medskip
\subsection{Non abelian zero-winding solutions}
The differential equations for the non-abelian version of the zero-winding solution can be
obtained by putting $n=0$ in our general equations. Actually, by taking $n=0$ in the
second equation in   (\ref{eqpsiyr}) one obtains the differential equation which determines
the dependence of
$r$ on the angle $\theta$, namely:
\beq
r_{\theta}\,=\,-{\cot\theta\over 2\coth2r}\,\,,
\eeq
which can be easily integrated, namely:
\beq
\sinh 2r\,=\,{C\over \sin\theta}\,\,.
\label{zerowindingr}
\eeq
Notice that, as in the abelian case, this solution has two spikes at $\theta=0,\pi$, where
$r$ diverges and, thus, the brane probe also extends infinitely in the radial direction.
Moreover, the minimal value of the radial coordinate, which we will denote by $r_*$, is
reached at $\theta=\pi/2$. This minimal value is related to the constant $C$ in eq. 
(\ref{zerowindingr}), namely $\sinh 2r_*=C$. 
It is readily verified  that for large $r$ this solution behaves exactly as the
zero-winding solution  in the abelian theory. Moreover, it follows from eq. (\ref{rel})
that, in this $n=0$ case, the angle $\tilde\varphi$ only depends on $\theta$. Actually,
the differential equations for the angles $\tilde\theta$, $\tilde\varphi$ and $\psi$ as
functions of $\theta$ are easily obtained from eqs. (\ref{eqtildetheta}), 
(\ref{eqtildevarphi}) and (\ref{rel}):
\bear
\partial_{\theta} \,\tilde\theta&=&-{\cos\psi\over \cosh 2r}\,\,,\rc\rc
\partial_{\theta} \,\tilde\varphi&=&-{1\over \cosh 2r}\,\,
{\sin\psi\over \sin\tilde\theta}\,\,,\rc\rc
\partial_{\theta} \,\psi&=&{\cot\tilde\theta\sin\psi\over\cosh 2r}\,\,.
\label{zerowindingangles}
\eear
By combining the equations of $\psi$ and $\tilde\theta$ one can easily get the
relation between these two angles, namely:
\beq
\sin\psi\,=\,{B\over \sin\tilde\theta}\,\,.
\label{psitildetheta}
\eeq
Notice that, for consistency, $B\le 1$ and $\sin\tilde\theta\ge B$. We can also
obtain $\tilde\varphi=\tilde\varphi(\tilde\theta)$ and 
$\tilde\theta=\tilde\theta(\theta)$:
\bear
&&\tilde\varphi=-\arctan\Bigg[\,
{B\cos\tilde\theta\over \sqrt{\sin^2\tilde\theta-B^2}}\Bigg]\,+\,
{\rm constant}\,\,,\rc\rc
&&-\arcsin\Bigg[\,{\cos\tilde\theta\over \sqrt{1-B^2}}\Bigg]=
\arcsin\Bigg[\,{\cos\theta\over \sqrt{1+C^2}}\Bigg]\,+\,
{\rm constant}\,\,.
\eear
Actually, much simpler equations for the embedding are obtained if one considers the
particular case in which the angle $\psi$ is constant. Notice that, as was pointed out
after eq. (\ref{eqpsiyr}), this only can happen if $\psi=0,\pi$ ($\mod \,\,2\pi$) (see
also the last equation in (\ref{zerowindingangles})). These solutions correspond to
taking the constant $B$ equal to zero in eq. (\ref{psitildetheta}).
Moreover, it follows from the eq. 
(\ref{zerowindingangles}) that $\tilde\varphi$ is an arbitrary constant in this case,
while the dependence of $\tilde\theta$ on $\theta$ can be obtained by combining eqs. 
(\ref{zerowindingr}) and (\ref{zerowindingangles}). If we denote $\cos\psi=\epsilon$,
with $\epsilon=\pm 1$, one has:
\beq
\sinh 2r\,=\,{\sinh 2r_*\over \sin\theta}\,\,,
\,\,\,\,\,\,\,\,\,\,\,\,\,\,\,\,\,\,\,
\sin (\,\tilde\theta\,-\,\tilde\theta_*\,)\,=\,\epsilon\,
{\cos\theta\over \cosh 2r_*}\,\,,
\label{zwprofile}
\eeq
where $\tilde\theta_*=\tilde\theta(\theta=\pi/2)$. Notice that there are four possible
values of $\psi$ in this zero-winding solution and, thus, the U(1) R-symmetry is broken
to $\ZZ_4$ in this case.

Let us finally point out that, also in this case, 
the  hamiltonian density ${\cal H}$ can be put as  ${\cal H}={\cal Z}+{\cal S}$,
where
${\cal S}$ is nonnegative (${\cal S}=0$ for the BPS solution) and 
${\cal Z}$ is given by:
\beq
{\cal Z}\,=\,-\partial_{\theta}\,\Bigg[\,
e^{2\phi}\,\cos\theta\bigg(\,r\,-\,{1\over 4}\,\coth 2r\,+\,
{r\over 2\sinh^22r}\,
\bigg)\,\Bigg]\,\,.
\eeq

\medskip
\subsection{Cylinder solutions}

We shall now show that there exists a general class of supersymmetric
embeddings for the non-abelian background. For convenience, let us
consider $r$ as worldvolume coordinate and let us assume that the
D5-brane is sitting at the north poles of the $(\theta,\varphi)$ and 
$(\tilde\theta,\tilde\varphi)$ two-spheres, \ie\ at 
$\theta=\tilde\theta=0$. In the remaining angular
coordinates  $\varphi$,  $\tilde\varphi$ and $\psi$,  the embedding  is characterized by
the equation:
\beq
{\varphi-\varphi_0\over p}\,=\,
{\tilde\varphi-\tilde\varphi_0\over q}\,=\,
{\psi-\psi_0\over s}\,\,,
\label{cylcurve}
\eeq
where $(\varphi_0,\tilde\varphi_0,  \psi_0)$ and $(p,q,s)$ are
constants. Notice that, if one of the constants of the denominator in
(\ref{cylcurve}) is zero, then the corresponding angle must be a
constant. Let us parametrize these embeddings by means of two worldvolume
coordinates $\sigma_1$ and $\sigma_2$, defined as follows:
\beq
\sigma_1\,=\,{\varphi-\varphi_0\over p}\,=\,
{\tilde\varphi-\tilde\varphi_0\over q}\,=\,
{\psi-\psi_0\over s}\,\,,
\,\,\,\,\,\,\,\,\,\,\,\,\,\,\,\,\,
\sigma_2\,=\,r\,\,.
\eeq
It is straightforward to demonstrate that the pullback to the
worldvolume of the forms $w^i$ and $A^i$ is given by:
\bear
&&P[\,w^1\,]\,=\,P[\,w^2\,]\,=\,0\,\,,
\,\,\,\,\,\,\,\,\,\,\,\,\,\,\,\,\,
P[\,w^3\,]\,=\,(q+s)\,d\sigma_1\,\,,\rc\rc
&&P[\,A^1\,]\,=\,P[\,A^2\,]\,=\,0\,\,,
\,\,\,\,\,\,\,\,\,\,\,\,\,\,\,\,\,
P[\,A^3\,]\,=\,-p\,d\sigma_1\,\,.
\eear
It follows from these results that the pullback of the frame
one-forms 
$e^i$ and $e^{\hat j}$  is zero for $i,j=1,2$, whereas $P[e^{\hat 3}]$
is non-vanishing. As a consequence, the induced Dirac matrices are:
\beq
\gamma_{\sigma_1}\,=\,{1\over 2}\,(p+q+s)\,e^{{\phi\over 4}}\,
\hat\Gamma_3\,\,,
\,\,\,\,\,\,\,\,\,\,\,\,\,\,\,\,\,\,\,\,\,\,\,\,\,\,\,\,\,\,\,\,\,\,
\gamma_{\sigma_2}\,=\,e^{{\phi\over 4}}\,\,\Gamma_r\,\,.
\eeq
The kappa symmetry matrix $\Gamma_{\kappa}$ for the
embedding at hand is:
\beq
\Gamma_{\kappa}\,=\,{e^{\phi}\over \sqrt{-g}}\,
\Gamma_{x^0\cdots x^3}\,\gamma_{\sigma_1\sigma_2}\,\,.
\eeq
Moreover, by using the projection conditions satisfied by the Killing
spinors $\epsilon$, one can prove that
\beq
\gamma_{\sigma_1\sigma_2}\,\epsilon\,=\,-
{p+q+s\over 2}\,\,e^{{\phi\over 2}}\,\,\Gamma_r\,\hat\Gamma_3\,
\epsilon\,=\,{p+q+s\over 2}\,\,e^{{\phi\over 2}}\,
\Big(\,\cos\alpha\,\Gamma_{12}\,+\,\sin\alpha\Gamma_1\hat\Gamma_2
\,\Big)\,\epsilon\,\,,
\eeq
and, since the determinant of the induced metric is
$\sqrt{-g}\,=\,e^{{3\phi\over 2}}\,{p+q+s\over 2}$,
it is immediate to
verify that the kappa symmetry projection 
$\Gamma_{\kappa}\epsilon=\epsilon$
coincides with the
projection  satisfied by the Killing spinors of the
background. Therefore, our brane probe preserves all the
supersymmetries of the background. Notice that the induced metric on
the worldvolume along the $\sigma_1,\sigma_2$ directions has the form
\beq
e^{{\phi\over 2}}\big[\,{(p+q+s)^2\over 4}\,
d\sigma_1^2\,+\,d\sigma_2^2
\,\big]\,\,,
\eeq
which is conformally equivalent to the metric of a cylinder. After a
simple calculation one can prove that the energy density of these
solutions is
\beq
{\cal H}\,=\,\partial_r\,\Bigg[\,e^{2\phi}\,
\bigg(\,pr\,+\,(q+s-p)\,\bigg(
{\coth2r\over 4}\,-\,
{r\over 2\sinh^22r}\,\bigg)\,
\bigg)\,\Bigg]\,\,.
\eeq
 One can also have cylinders  located at the south pole of the
$(\theta, \varphi)$ and $(\tilde\theta, \tilde\varphi)$ two-spheres.
Indeed, the above equations remain valid if $\theta=\pi$
($\tilde\theta=\pi$) if one changes $p\rightarrow -p$ 
($q\rightarrow -q$, respectively). On the other hand, if $p=0$ the
angle $\varphi$ is constant and, as the pullback of the $e^{i}$ frame
one-forms also vanishes when $\theta$ is also constant, it follows
that $\theta$ can have any constant value when $p=0$. Similarly, if
$q=0$ one necessarily has $\tilde\varphi=\tilde\varphi_0$ and 
$\tilde\theta$ can be an arbitrary constant in this case.  

When $p=1$, $q=n$ and $s=2m$, the angular part of the embedding is the
same as in the $(n,m)$-winding solitons. Actually, these cylinder
solutions correspond to formally taking $r_*\rightarrow-\infty$ is the
abelian solution of eq. (\ref{nmsol}). This forces one to take
$\theta=0,\pi$ and, thus one can regard the cylinder as a zoom which
magnifies   the region in which the probe goes to infinity. One can
also get cylinder embeddings by consider the limit of the non-abelian
solutions in which the probe reaches the point $r=0$. For example, by taking $r_*=0$ in
eq. (\ref{noabflavor}) one gets the $p=q=1$, $s=0$ cylinder solutions while the 
$r_*\to 0$ limit of the embedding (\ref{zwprofile}) corresponds to a cylinder with $p=1$
and  $q=s=0$. Actually, when one takes the $r_*\to 0$ limit of these non-abelian
embeddings one obtains two cylinder solutions, one with $\theta=0$ and the other with
$\theta=\pi$. This suggests that, in order to obtain a consistent solution, one must
combine in general two cylinders located at each of the two poles of the
$(\theta,\varphi)$ two-sphere. Notice that this is also required if one imposes the
condition of RR charge neutrality of the two-sphere at infinity.

\setcounter{equation}{0}
\section{Quadratic fluctuations around the unit-winding embedding}
\medskip
As mentioned in the introduction, we are now going to consider some of the brane probe
configurations previously found as flavor branes, which will allow us to introduce 
dynamical quarks in the ${\cal N}=1$ SYM theory. Following refs. \cite{KK, KKW, KMMW},
the spectrum of quadratic fluctuations of the brane probe will be interpreted as the meson
spectrum of ${\cal N}=1$ SQCD. So, let us try to elaborate on the reasons to consider 
these probes as the addition of flavors to the field theory dual.
In fact, when considering the  't Hooft expansion for large number of colors, 
the role of flavors is played by the boundaries of the Feynman graph. From 
a gravity perspective, these boundaries correspond to the addition of D-branes and open
strings in the game.

In our case, we have a system of $N$ D5-branes wrapping a two-cycle inside 
the 
resolved conifold and $N_f$ D5-branes that wrap another two-submanifold, thus 
introducing $N_f$ flavors in the $SU(N) $ gauge theory. Taking the 
decoupling limit with $g_s \alpha' N$ fixed and large is equivalent to replacing 
the $N$ D5-branes by the geometry they generate (the one studied in 
section 2), while the $N_f$-D5 branes that do not backreact (because we 
take $N_f$ much smaller than $N$) are treated as 
probes. From a gauge theory perspective, this is equivalent to consider 
the dynamics of gluons and gluinos coupled to  fundamentals, but 
neglecting the backreaction of the latter. Of course it would be of 
great interest to find the backreacted solution.

The way of adding fundamental fields in this gauge theory from a string 
theory perspective was discussed in \cite{WangHu}, where two possible 
ways, adding $D9$ branes or adding $D5$
branes as probes, were proposed. In this paper we are considering the  cleaner case of 
$D5$ probes. A careful analysis 
of the open string spectrum shows the existence of a four dimensional 
gauge ${\cal N}=1$ vector multiplet and a complex scalar multiplet. This 
is 
the spectrum of SQCD. In the case analyzed below we will consider abelian 
DBI actions for the probes, so that we will be dealing with the 
 $N_f=1$ case.

 We have found several brane configurations
in the non-abelian background which, in principle,  could be suitable to generate  the
meson spectrum. 
One of the requirements we should demand to these configurations is that
they must incorporate some scale parameter which could be used to generate the mass scale
of the quarks. Within our framework such a mass scale is nothing but the minimal distance
between the flavor brane and the origin, \ie\ what we have  denoted by $r_*$. This
requirement allows to discard the cylinder solutions we have found since they reach the
origin and have no such a mass scale. We are thus left with the unit-winding solutions of
section 6.1 and the zero-winding solutions of section  6.2 as the only analytical
solutions we have found for the non-abelian background.

In this section we shall analyze the fluctuations around the unit-winding solutions of
section 6.1. We have several reasons for this election. First of all, the unperturbed
unit-winding embedding is simpler. Secondly, we will show in appendix B that the UV
behaviour of the  fluctuations is better in the unit-winding configuration than in the
zero-winding embedding. Thirdly,  the unit-winding embeddings of constant $\psi$
incorporate the correct pattern ${\rm U(1)}\to \ZZ_2$ of R-symmetry breaking, whereas
for the zero-winding embeddings of eq. (\ref{zwprofile}) the U(1) symmetry is broken to a
$\ZZ_4$ subgroup.

Recall from section 6.1 that we have two possible solutions with
$\psi=0,\pi\,\,(\mod\,2\pi)$, which are the ones displayed in eqs. (\ref{noabflavor})
and (\ref{nabcosh}). As discussed in section 6.1, the solution of eq. (\ref{nabcosh})
contains a one-parameter subfamily of embeddings which reach the origin and, thus, they
should correspond to massless dynamical quarks. On the contrary,  the embeddings of
eq. (\ref{noabflavor}) pass through the origin only in one case, \ie\ when $r_*=0$ and,
somehow, the limit in which the quarks are massless is uniquely defined. Recall that for
$r_*=0$ the solution (\ref{noabflavor})  is identical to the unit-winding cylinder. For
these reasons we consider the configuration displayed in eq.  (\ref{noabflavor}) more
adequate for our purposes and we will use it as the unperturbed flavor brane.

 We will consider first in subsection 7.1  the
fluctuations of the scalar transverse to the transverse probe, while in subsection  7.2
we will study the fluctuations of the worldvolume gauge field.  The gauge theory
interpretation of the results will be discussed in subsection 7.3

\subsection{Scalar mesons}
\medskip
Let us consider a non-abelian unit-winding  embedding with  
$\tilde\theta=\theta$, $\tilde\varphi=\varphi+{\rm constant}$ and $\psi=\pi \,(\mod\,
2\pi)$. For convenience we take first $r$ as worldvolume coordinate and consider
$\theta$ as a function of $r$, $\varphi$  and of the unwrapped coordinates $x$, \ie\
$\theta=\theta(r,x,\varphi)$.  The lagrangian density for such embedding can be easily
obtained by computing the induced metric. One gets:
\bear
&&{\cal L}=-e^{2\phi}\,\sin\theta\,\times\rc\rc
&&\times\Bigg[\,
\sqrt{\bigg[\,1+r\tanh r\Big(\,(\partial_r\theta)^2+
(\partial_x\theta)^2\,+\, {1\over \cos^2\theta+r\coth r\sin^2\theta}\,
(\partial_{\varphi}\theta)^2\,\Big)
\bigg]\,
\big(r\coth r+\cot^2\theta\big)}\,+\,\rc\rc
&&\,\,\,\,\,\,\,\,\,\,\,\,\,\,\,\,\,\,\,\,\,\,\,\,\,\,\,\,\,\,\,\,
+\,r\partial_r\theta-\cot\theta\,\Bigg]\,\,,
\eear
where we have neglected the term $\partial_r (r\cos\theta e^{2\phi})$ which,
being a total radial derivative, does not contribute to the equations of
motion. 

We are going to expand this lagrangian around the corresponding non-abelian unit-winding
configuration obtained in section 6.1. Actually, by taking in eq. (\ref{noabflavor})
$\eta=+1$  one  obtains a  configuration with $\psi=\pi \,\,(\mod \,2\pi)$, 
which corresponds to a function $\theta=\theta_0(r)$, given
by:
\beq
\sin\theta_0(r)={\sinh r_{*}\over \sinh r}\,\,,
\label{thetazero}
\eeq
where $r_{*}$ is the minimum value of $r$ and $r_{*}\le r<\infty$. It is clear from this
equation that with the coordinate $r$ we can only describe one-half of the brane probe:
the one that is wrapped, say, on the north hemisphere of the two-sphere, in which
$\theta\in (0,\pi/2)$. Outside this interval $\theta_0(r)$ is a double-valued function of
$r$. Let us put
\beq
\theta(r,x,\varphi)\,=\,\theta_0(r)\,+\,\chi(r,x,\varphi)\,\,,
\eeq
and expand ${\cal L}$ up to quadratic order in $\chi$. Using the first-order
equation satisfied by $\theta_0(r)$, namely:
\beq
\partial_r\theta_0\,=\,-\coth r\tan\theta_0\,\,,
\eeq
we get
\bear
{\cal L}&=&-{1\over 2}\,\,
{e^{2\phi}\over 1+r\coth r\tan^2\theta_0}\,\,\Bigg[\,\,
r\tanh r\cos\theta_0\,(\partial_r\chi)^2\,+\,
{2r\over \cos\theta_0}\,\chi\,\partial_r\chi\,+\,
{r\coth r\over \cos^3\theta_0}\,\,\chi^2\,\,\Bigg]\,-\rc\rc
&&-\,{1\over 2}\,\,e^{2\phi}\,r\tanh r\,\Big[\,
\cos\theta_0\, (\partial_x\chi)^2\,+\,
{1\over \cos\theta_0\,(\,1+r\coth r\tan^2\theta_0\,)}\,\,(\partial_{\varphi}\chi)^2
\,\,\Big]\,\,.
\label{lagrafluct}
\eear

In the equations of motion derived from this lagrangian we will perform the
ansatz:
\beq
\chi(r,x,\varphi)=e^{ikx}\,e^{il\varphi}\,\,\xi(r)\,\,,
\label{chiansatz}
\eeq
where, as $\varphi$ is a periodically identified coordinate,  $l$ must be an integer and
$k$ is a four-vector whose square determines the four-dimensional mass $M$ of the
fluctuation mode:
\beq
M^2=-k^2\,\,.
\label{masssquare}
\eeq
By substituting the functions (\ref{chiansatz}) in the equation of motion that follows
from the lagrangian (\ref{lagrafluct}), one gets a second order differential equation
which is rather complicated and that can only be solved numerically. However, it is not
difficult to obtain analytically the asymptotic behaviour of $\xi(r)$. This has been done
in appendix B and we will now use these results to explore the nature of the
fluctuations. For large $r$, \ie\ in the UV, one gets (see eq. (\ref{UVasymp})) that
$\xi(r)$ vanishes exponentially in the form:
\beq
\xi(r)\,\sim\,{e^{-r}\over r^{{1\over 4}}}
\cos \big[\sqrt{M^2-l^2}\,\,r\,+\,\delta\,\big]\,\,,
\,\,\,\,\,\,\,\,\,\,\,\,\,\,\,\,\,
(r\rightarrow\infty)\,\,,
\label{UVasympmain}
\eeq
where $\delta$ is a phase and we are assuming that $M^2\ge l^2$. 
For $M^2< l^2$ the fluctuations do not oscillate in the UV and we will not be able to
impose the appropriate boundary conditions (see below).
Notice that our
unperturbed solution $\theta_0(r)$ also decreases in the UV as
\beq
\theta_0(r)\sim e^{-r}\,\,
\,\,\,\,\,\,\,\,\,\,\,\,\,\,\,\,\,
(r\rightarrow\infty)\,\,.
\eeq
Thus $\xi(r)/\theta_0(r)\rightarrow 0$ as $r\rightarrow\infty$ and the first-order
expansion we are performing continues\footnote{For the zero-winding solution, on the
contrary, the ratio $\xi(r)/\theta_0(r)$ diverges in the UV (see appendix B).}
to be  valid in the UV. On the other hand,
for
$r$ close to $r_*$ there are two independent solutions, one of them is finite at 
 $r=r_*$ while the other diverges as
\beq
\xi(r)\,\sim\,{1\over \sqrt{r-r_*}}\,\,.
\label{xiUV}
\eeq

Let us now see how one can use the information on the asymptotic behaviour of the
fluctuation modes to extract their value for the full range of the radial
coordinate. First of all, it is clear  that, in principle,  by consistency with the
type of expansion we are  adopting, one should require that $\xi<<\theta_0$. Thus,
one should discard the solutions which diverge in the infrared (see, however, the
discussion below).
Moreover, the behaviour of
the fluctuations $\xi$ for large $r$ should be determined by some normalizability
conditions. The corresponding norm would be an expression of the form:
\beq
\int_{0}^{\infty}\,dr\sqrt{\gamma}\,\,\,\xi^2\,\,,
\eeq
where $\sqrt{\gamma}$ is some measure, which can be determined by looking at
the lagrangian (\ref{lagrafluct}). If we regard $\chi$ as a scalar field with the
standard normalization in a curved space, then $\sqrt{\gamma}$ is just the
coefficient of the kinetic term ${1\over 2}\,(\partial_r\chi)^2$ in ${\cal L}$,
namely
\beq
\sqrt{\gamma}\,=\,e^{2\phi}\,\,
{r\tanh r \cos\theta_0\over 1+r\coth r \tan^2\theta_0}\,\,.
\label{measure}
\eeq
For large $r$, $\sqrt{\gamma}$  behaves as
\beq
\sqrt{\gamma} \,\,(\,\,r\rightarrow\infty\,\,)\,\approx\,r^{{1\over 2}}\,
e^{2r}\,\,.
\label{UVmeasure}
\eeq
Notice that the factors on the right-hand side of eq. (\ref{UVmeasure}) cancel
against the exponentials and power factors of $\xi^2$ in the UV 
(see eq. (\ref{UVasympmain})).
As a consequence, all solutions have infinite norm.

The reason for the bad behaviour we have just discovered is the exponential
blow up of the dilaton in the UV which invalidates the supergravity
approximation. Actually, if one wishes to push the theory to the UV one has to
perform an S-duality, which basically changes $e^{2\phi}$ by 
$e^{-2\phi}$. The S-dual theory corresponds to wrapped Neveu-Schwarz fivebranes
and is the supergravity dual of a little string theory. Notice that, by
changing  $e^{2\phi}\rightarrow e^{-2\phi} $ in the measure (\ref{measure}), 
all solutions
become normalizable, which is  as bad as having no normalizable solutions at
all. Moreover, it is unclear how to perform an S-duality in our D5-brane probe
and convert it in a Neveu-Schwarz fivebrane for large values of the radial
coordinate. 

A problem similar to the one we are facing here appeared in ref. \cite{Pons} in the
calculation of the glueball spectrum for this background. It was argued in
this reference that, in order to have a discrete spectrum, one has to introduce a
cut-off to discriminate between the two regimes of the theory. Notice that,
since they extend infinitely in the radial direction, we cannot avoid that our
D5-brane probe explores the deep UV region. However, what we can do is to
consider fluctuations that are significantly nonzero
only on scales in which one can safely trust the supergravity approximation. 
In ref. \cite{Pons} it was proposed to implement this condition by requiring the
fluctuation to vanish at some conveniently chosen UV cut-off $\Lambda$.
Translated to our situation, this proposal amounts to requiring:
\beq
\xi(r)\big|_{r=\Lambda}\,=\,0\,\,.
\label{cutoff}
\eeq
This condition, together with the regularity of $\xi(r)$ at $r=r_{*}$, produces
a discrete spectrum which we shall explore below. Notice that, for consistency
with the general picture described above, in addition to having a node at
$r=\Lambda$ as in eq. (\ref{cutoff}), the function $\xi(r)$ should be small for
$r$ close to the UV cut-off. This condition can be fulfilled by adjusting
appropriately the mass scale of our solution, \ie\ the minimal distance $r_*$
between the probe and the origin, in such a way that 
$r_*$ is not too close to $\Lambda$. 

Notice that, by imposing the boundary condition (\ref{cutoff}) on the
fluctuations, we are effectively introducing an infinite wall located at the
UV cut-off. The introduction of this wall allows to have a discrete spectrum
and should be regarded as a physical condition which implements the correct
range of validity of the background geometry as a supergravity dual of 
${\cal N}=1$  Yang-Mills. Even if this regularisation could appear too rude and
unnatural, the results obtained by using it for the first glueball masses are
not too bad \cite{Pons}.

The cut-off scale $\Lambda$ should not be a new scale but instead it should 
be obtainable from the background geometry itself. The proposal of ref. \cite{Pons} is
to take $\Lambda$ as the scale of gaugino condensation, which is believed to
correspond to the point at which the function $a(r)$ approaches its asymptotic
value $a=0$. A more pragmatic point of view, to which we will adhere here, is
just taking the value of $\Lambda$ which gives reasonable values for the
glueball masses. In ref. \cite{Pons} the value $\Lambda=2$ was needed to fit the
glueball masses obtained from lattice calculations, whereas with 
$\Lambda=3.5$ one gets a glueball spectrum which resembles that predicted by
other supergravity models. Notice that from $r=0$ to $r=3$ the effective string
coupling constant $e^{\phi}$ increases in an order of magnitude. From our point
of view it is also natural to look at the effect of the background on our brane
probe. In this sense it is interesting to point out that for $r_*=2-3$ onwards
the abelian and non-abelian embeddings are indistinguishable (see sect. 6.1). 
\begin{figure}
\centerline{\hskip -.8in \epsffile{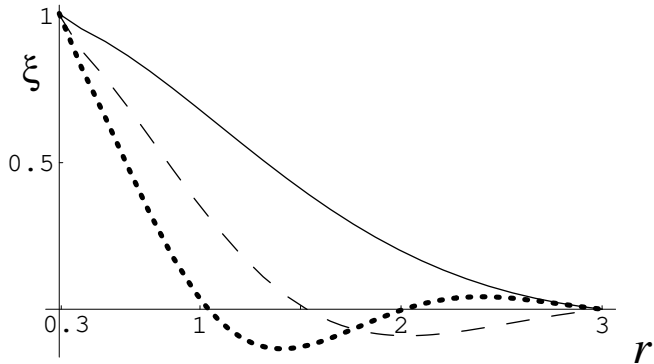}}
\caption{Graphic representation of the first three fluctuation modes for $r_*=0.3$, 
$\Lambda=3$ and $l=0$. The three curves have been normalized to have $\xi(r_*)=1$. }
\label{fig4}
\end{figure}

We have performed the numerical integration of the equation of motion of $\xi(r)$
subject to the boundary condition (\ref{cutoff}) by means of the shooting technique.
For a generic value of the mass $M$ the solution diverges at $r=r_*$. Only for some
discrete set of masses the fluctuations are regular in the IR. In figure 4 we show the
first three modes obtained by this procedure for $r_*=0.3$ and $\Lambda=3$. From this
figure,  we notice that the number of zeroes of $\xi(r)$ grows with the mass. In general
one observes that the $n^{th}$ mode has $n-1$ nodes in the region $r_*<r<\Lambda$, in
agreement with the general expectation for this type of boundary value problems.
Moreover, for $l=0$, the mass $M$ grows linearly with the number of nodes (see below).

At this point it is interesting to pause a while and discuss the suitability of our
election of $r$ as worldvolume coordinate. Although this coordinate is certainly very
useful to extract the asymptotic behavior of the fluctuations (specially in the UV), we
should keep in mind that we are only describing one half of the brane, \ie\ the one
corresponding to one of the two hemispheres of the two-sphere. On the other hand, the
election of the angle as excited scalar has some subtleties which we now discuss.
Actually, to describe the displacement of  the brane probe with respect to its
unperturbed configuration it is physically more sensible to use the coordinate $y$,
defined in eq. (\ref{cartesian}). Accordingly, let us define the function $y(r,x,\varphi)$
as
\beq
y(r,x,\varphi)\,=\,r\sin\theta(r,x,\varphi)\,\,.
\eeq
Let us put in this equation $\theta(r,x,\varphi)=\theta_0(r)\,+\,\chi(r,x,\varphi)$. At
the  linear order in 
$\chi$ we are working, $y$ can be written as
\beq
y(r,x,\varphi)\,=\,y_0(r)\,+\,r\cos\theta_0(r)\,\chi(r,x,\varphi)\,\,,
\label{yexpanded}
\eeq
where $y_0(r)\equiv r\sin\theta_0(r)$ corresponds to the unperturbed brane. Notice, first
of all, that for those modes in which $\chi(r_*,x,\varphi)$ is finite, the 
fluctuation term in $y(r_*,x,\varphi)$  vanishes since $\cos\theta_0(r)\rightarrow 0$
when 
$r\rightarrow r_*$. Then
\beq
y(r_*,x,\varphi)\,=\,y_0(r_*)\,=\,r_*\,\,\,\,
{\rm if}\,\,\, \chi(r_*,x,\varphi)\,\,\,\,{\rm is\,\,\,finite}\,\,.
\eeq
Thus, by considering those modes $\chi$ that are regular at $r= r_*$ we are
effectively restricting ourselves to the modes which have a node in the $y$ coordinate at 
$r= r_*$. If, on the contrary $\chi$ diverges for $r\approx r_*$, we know from eq.
(\ref{xiUV}) that it behaves as $\chi\,\approx\,1/ \sqrt{r-r_*}$. 
But we also know that $\cos\theta_0(r)\rightarrow 0$
when $r\rightarrow r_*$ and, in fact, the second term on the right-hand side of eq. 
(\ref{yexpanded}) remains undetermined. The precise form in which 
$\cos\theta_0(r_*)$ vanishes can be read from eq. (\ref{sincos}), namely
$\cos\theta_0(r)\,\approx\,\sqrt{r-r_*}$ for $r\approx r_*$. 
Therefore, even if $\chi$ diverges at $r=r_*$, we could have, in the linearized
approximation  we are adopting, a finite value for $y(r_*,x,\varphi)$. Thus, modes with
$\chi(r_*,x,\varphi)\rightarrow\infty$ should not be discarded. 
Actually, $y(r_*,x,\varphi)-y_0(r_*)$, although finite, is undetermined  in eq.
(\ref{yexpanded}) for these modes and, in order to obtain its allowed values we should
impose a boundary condition at the other half of the brane. As previously mentioned, this
cannot be done when $r$ is taken as worldvolume coordinate. Therefore, it is convenient
to come back to the formalism of sects. 3-5, in which
$\theta$ had been chosen as one of the worldvolume coordinates. In this approach, the
unperturbed brane configuration is described by a function $r_0(\theta)$, given by:
\beq
\sinh r_0(\theta)\,=\,{\sinh r_*\over \sin\theta}\,\,,
\eeq
and the brane embedding is characterized by a function $r=r(\theta, x,\varphi)$, which
we expand around $r_0(\theta)$ as follows:
\beq
r(\theta, x,\varphi)\,=\,r_0(\theta)\,+\,\rho(\theta, x, \varphi)\,\,.
\eeq
Plugging this expansion into the DBI lagrangian of eq. (\ref{DBI}) and keeping up to
second order terms in $\rho$, one gets the following lagrangian density:
\bear
{\cal L}&=&-{1\over 2}\,\,
{e^{2\phi}\sin\theta\,\, r_0\over r_0+\cot^2\theta\tanh r_0}
\Bigg[\coth r_0\,\big(\,\partial_{\theta}\,\rho\,\big)^2\,+\,
{2\cot\theta\over \sinh r_0\cosh r_0}\,\,\rho \partial_{\theta}\,\rho+
{\cot^2\theta\over \sinh r_0 \cosh^3 r_0}\,\,\rho^2\,\,\Bigg]\,-\rc\rc
&&\,\,-\,\,
{e^{2\phi}\sin\theta\,\, r_0\over 2} \,\Bigg[\,\big(\partial_{x}\,\rho\,\big)^2
\,+\,{1\over \cos^2\theta\,\big(1+r_0\coth r_0\tan^2\theta\,\big)}\,\,
\big(\partial_{\varphi}\,\rho\,\big)^2\,\Bigg]\,\,.
\label{Lfluct}
\eear
Similarly to what we have done with the lagrangian (\ref{lagrafluct}), we will look for
solutions of the equations of motion of ${\cal L}$ which have the form:
\beq
\rho(\theta, x,\varphi)\,=\,e^{ikx}\,e^{il\varphi}\,\,\zeta (\theta)\,\,,
\eeq
where $l$ is an integer and $k^2=-M^2$. As before, in order to get a discrete spectrum
one must impose some boundary conditions. In the present approach we should cutoff the
regions close to the two poles of the two-sphere. Accordingly, let us define the
following angle
\beq
\sin\theta_{\Lambda}\equiv {\sinh r_*\over \sinh \Lambda}\,\,,
\,\,\,\,\,\,\,\,\,\,\,\,\,\,\,
\theta_{\Lambda}\in (0, {\pi\over 2})\,\,,
\eeq
where, as indicated, we are taking $\theta_{\Lambda}$ in the range 
$0<\theta_{\Lambda}<\pi/2$. Notice that $\theta_{\Lambda}\rightarrow 0$ if 
$\Lambda\rightarrow\infty$ as it should. Clearly $\theta_{\Lambda}$ and
$\pi-\theta_{\Lambda}$ are the two angles that correspond to the radial scale $\Lambda$.
Therefore, we impose the following boundary conditions to our fluctuation
\beq
\zeta(\theta_{\Lambda})\,=\,\zeta(\pi-\theta_{\Lambda})\,=\,0\,\,.
\label{bdytheta}
\eeq
\begin{figure}
\centerline{\hskip -.8in \epsffile{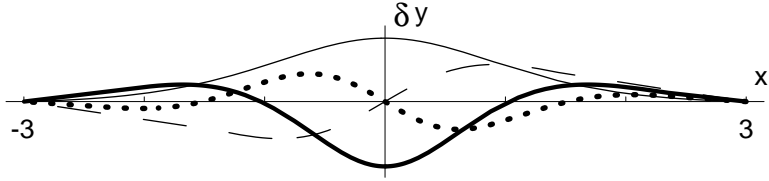}}
\caption{Plot of $\delta y\equiv\zeta\sin\theta$ versus $x= r\cos\theta$ for the first
four modes for $r_*=0.3$, $\Lambda=3$ and $l=0$. The dashed and dotted curves pass through
the origin and correspond to the first two modes of figure 3.}
\label{fig5}
\end{figure}

The equations of motion derived from (\ref{Lfluct}), subjected to the boundary conditions
(\ref{bdytheta}) can be integrated numerically by means of the shooting technique. One
first enforces the condition at $\theta=\theta_{\Lambda}$ and then varies the mass $M$
until $\zeta(\pi-\theta_{\Lambda})$ vanishes. This only happens for a discrete set of
values of the mass $M$. For a given value of $l$, let us order the solutions in
increasing value of the mass. In general one notices that the $n^{th}$ mode has $n-1$
nodes in the interval $\theta_{\Lambda}<\theta<\pi-\theta_{\Lambda}$ and for $n$ even
(odd) the function $\zeta(\theta)$ is odd (even) under 
$\theta\rightarrow \pi-\theta$. In figure 5 we have plotted the first four modes
corresponding to $r_*=0.3$, $\Lambda=3$ and $l=0$. The modes odd under 
$\theta\rightarrow \pi-\theta$ vanish at $\theta=\pi/2$ and their masses and shapes match
those found with the lagrangian (\ref{lagrafluct}) and the boundary condition
(\ref{cutoff}). On the contrary, the modes with an even number of nodes in 
$\theta_{\Lambda}<\theta<\pi-\theta_{\Lambda}$ are the ones we were missing in the
formulation in which $r$ is taken as worldvolume coordinate.

\begin{figure}
\centerline{\hskip -.8in \epsffile{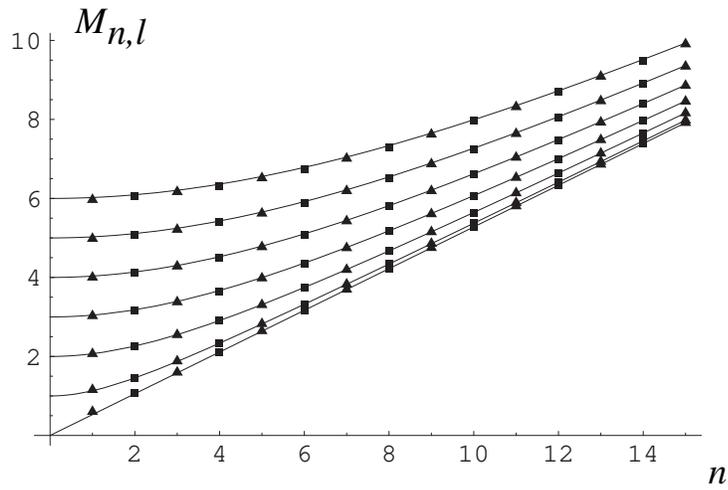}}
\caption{Mass spectrum for $r_*=0.3$ and $\Lambda=3$ for $l=0,\cdots, 6$. The solid
lines correspond to the right-hand side of eq. (\ref{spectrum}). The triangles (squares)
are the masses of the modes $\zeta(\theta)$ which are even (odd) under
$\theta\to\pi-\theta$.}
\label{fig6}
\end{figure}

Let $M_{n,l}(r_*,\Lambda)$ be the mass corresponding to the $n^{th}$ mode for a given
value $l$ and the mass scales $r_*$ and $\Lambda$. Our numerical results are compatible
with an expression of $M_{n,l}(r_*,\Lambda)$ of the form:
\beq
M_{n,l}(r_*,\Lambda)\,=\,\sqrt{m^2(r_*,\Lambda)\,\,n^2+l^2}
\label{spectrum}
\eeq
To illustrate this fact we have plotted in figure 6 the values of $M_{n,l}(r_*,\Lambda)$
for $r_*=0.3$ and $\Lambda=3$, together with the curves corresponding to the right-hand
side of (\ref{spectrum}).

\begin{figure}
\centerline{\hskip -.8in \epsffile{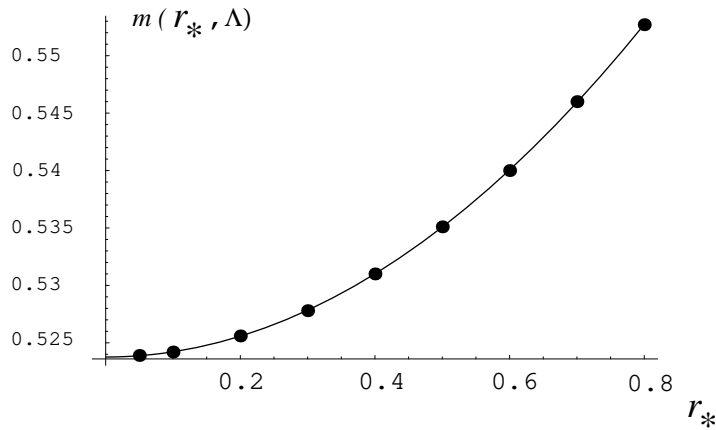}}
\caption{Dependence of $m(r_*,\Lambda)$ on $r_*$ for $\Lambda=3$. The solid line is a fit
to the quadratic function (\ref{parabola}).}
\label{fig7}
\end{figure}

We have also studied the dependence of the coefficient $m(r_*,\Lambda)$ on the two scales
$(r_*,\Lambda)$. Recall that $r_*$ is the minimal separation of the brane probe from the
origin and, thus, can be naturally identified with the mass of the quarks. 
We obtained that   $m(r_*,\Lambda)$ can be represented by an expression of the type:
\beq
m(r_*,\Lambda)\,=\,{\pi\over 2\Lambda}\,+\,b(\Lambda)\,r^{2}_*\,\,.
\label{parabola}
\eeq
The coefficients on the right-hand side of eq. (\ref{parabola}) have been obtained by a
fit of $m(r_*,\Lambda)$ to a quadratic expression in $r_*$. An example of this fit is presented
in figure 7. The first term in eq. (\ref{parabola}) is a universal term (independent of 
$r_*$) which can be regarded as a finite size effect induced by our regularisation
procedure. We also have determined the dependence of the coefficient $b$ on $\Lambda$ and
it turns out that one can fit $b(\Lambda)$ to the expression
$b(\Lambda)=0.23\,\Lambda^{-2}\,+\,0.53\,\Lambda^{-3}$

We have obtained a very regular mass spectrum of 
particles, classified by two quantum numbers $n,l$ (see eqs. (\ref{spectrum}) and 
(\ref{parabola})). We can offer an interpretation of these formulas. Indeed, recall that
the two-submanifold on which we are wrapping the brane
is topologically like a cylinder, with the compact direction parametrized by
the angle $\varphi$. The quantum number $l$ is precisely the eigenvalue of the operator
$-i\partial_{\varphi}$, which generates the shifts of $\varphi$ and, indeed, the
dependence on $l$ of the mass displayed in eq. (\ref{spectrum}) is the typical one for a
Kaluza-Klein reduction along a compact coordinate. Therefore, we should interpret the
mesons with $l\not=0$ as composed by Kaluza-Klein modes.
 Moreover,
the term in (\ref{parabola}) proportional to 
$r_*^2$ can be understood as the contribution coming from the mass of the 
constituent quarks, while the term $\frac{\pi}{2 \Lambda}$ can be 
interpreted as a contribution coming from the `finite size` of the meson. 
Indeed, 	it looks like a  Casimir energy and it is originated in the 
presence of the cut-off $\Lambda$.

It is perhaps convenient to emphasize again that the cut-off 
procedure implemented here is a very natural procedure for this 
type of computations in this 
supergravity dual. In fact, given that the mesons are an IR effect in 
SQCD, we expect  the contributions of high energy effects  to be 
irrelevant or negligible in the physical properties of the meson itself.
This is indeed what we are doing when cutting off the integration range. 
We are just taking into account the `non-abelian` part of the probes, 
while neglecting the `abelian` part or, equivalently, discarding high energy 
contributions.

\subsection{Vector mesons}
\medskip

Let us now excite a gauge field that is living in the worldvolume of the brane. The
linearized equations of motion are:
\beq
\partial_{m}\,\Big[\,e^{-\phi}\,\sqrt{-g_{st}}\,{\cal F}^{nm}\,\Big]\,=\,0\,\,,
\label{vectoreom}
\eeq
where $g_{st}$ is the determinant of the induced metric in the string frame
and ${\cal F}^{nm}$ is the field strength of the worldvolume gauge field ${\cal A}_m$,
\ie\ ${\cal F}_{nm}=\partial_n{\cal A}_m-\partial_m{\cal A}_n$. 
Let us assume that the only non-vanishing components of the gauge field ${\cal A}$ are
those along the unwrapped directions $x^{\mu}$ of the worldvolume of the brane. 
In what follows we are going to use $\theta$ as 
worldvolume coordinate. Let us put
\beq
{\cal A}_{\mu} (\theta,x,\varphi)\,=\,\epsilon_{\mu}\varsigma (\theta)\,\,
e^{ikx}\,\,e^{il\varphi}\,\,,
\eeq
where $\epsilon_{\mu}$ is a constant polarization four-vector and $l$ must be an integer.
It follows from the equations of motion (\ref{vectoreom}) that $\epsilon_{\mu}$ must be
transverse,
\ie:
\beq
k^{\mu}\,\epsilon_{\mu}\,=\,0\,\,,
\eeq
and that $\varsigma (\theta)$ must satisfy the following second-order differential
equation:
\beq
\partial_{\theta}\,\Big[\,{\sin\theta\over \tanh r_0}\,
\partial_{\theta}\,\varsigma\,\Big]\,+\,
{\tanh r_0\over \sin\theta}\,\,\Bigg[\,M^2
\,\,\bigg(\cos^2\theta\,+\,r_0\,\coth r_0\,\sin^2\theta\bigg)\,-\,l^2\,\Bigg]
\varsigma\,=\,0\,\,,
\label{vectorfluct}
\eeq
where, as in eq. (\ref{masssquare}), $k^2=-M^2$.
We have solved numerically this equation by means of the shooting technique with the
boundary conditions
\beq
\varsigma(\theta_{\Lambda})\,=\,\varsigma(\pi-\theta_{\Lambda})\,=\,0\,\,.
\eeq
Surprisingly, the set of possible values of $M$ is given by the same expression as in 
the scalar meson case, \ie\ eq. (\ref{spectrum}), with a coefficient function
$m(r_*,\Lambda)$ which is, within the accuracy of our approximate calculation, equal
numerically to that of the scalar mesons. This is quite remarkable since the differential
equations we are solving in both cases are quite different (the equation satisfied by
$\zeta(\theta)$ is much more complicated than eq. (\ref{vectorfluct})). Thus, to
summarize, we predict a degeneracy between the scalar and vector mesons in the
corresponding ${\cal N}=1$ SYM theory.

\subsection{Gauge theory interpretation}
\medskip

Let us comment on some gauge theory aspects that can be read from the 
brane probes in sections 3 to 7.
For this purpose, we shall concentrate on the main solutions that have been 
used, namely, the abelian  solution with unit winding 
($n=1$) in eq. (\ref{BM}) and its non-abelian extension of eq. 
(\ref{noabflavor}).

First of all, let us analyze the R-symmetry of the gauge theory from the probe
viewpoint. It is clear that the abelian solution (\ref{BM}) is invariant under 
shifts of $\psi$ by a constant since the value of this angle is an arbitrary constant in
this solution. This symmetry has been identified as the geometric dual of the R-symmetry
\cite{MN,KOW,GHP}. 
Actually, this is  not a $ U(1)$ invariance of the background because
\cite{MN,KOW,GHP} of the presence of the RR  form that selects, by consistency, only some
particular values of the angle
$\psi$, \ie\ $\psi = \frac{2 \pi n}{N}$ with $1\le n\le 2N$. So, the abelian probes can
see a 
$\ZZ_{2N}$ symmetry. In contrast, when we consider the non-abelian probe, the solution 
(\ref{noabflavor}) selects two particular values of  the angle $\psi$, thus breaking
$\ZZ_{2N}$ down  to  $\ZZ_2$. Notice that the $U(1)\to \ZZ_{2N}$ breaking is an UV effect
(it takes place already in the abelian background) while the 
 $\ZZ_{2N}\to \ZZ_2$ breaking is an IR effect which appears only when one considers the
full non-abelian regular background.
This same  breaking pattern was observed in the case of SQCD with massive flavors.
Indeed,  as showed in \cite{Affleck:1983mk}, 
the theory with massive flavors has a non-anomalous discrete  
$\ZZ_{2N}$ R-symmetry, given by 
(component fields are used, $\lambda$ is the gaugino and $\Phi$  and $\bar \Phi$ are the 
squarks):

\beq
\lambda \to e^{- i \pi n/N}\lambda\,\,, ~
\,\,\,\,\,\,\,\, \Phi \to e^{- i \pi n/ N} \Phi\,\,,
\,\,\,\,\,\,\,\, \bar\Phi \to e^{- i \pi n/ N} \bar\Phi\,\,,
\eeq
with $n= 1, ...2N$

As shown in 
\cite{Affleck:1983mk}, this $\ZZ_{2N}$ symmetry is broken down to $\ZZ_2$ by the formation of a
squark condensate.
Indeed, one can see that $<\Phi \bar{\Phi}>$ transforms as 
$<\Phi \bar{\Phi}> \to e^{-2i\pi n/N}<\Phi \bar{\Phi}>$ leaving us with a $\ZZ_2$.
Besides, the squark condensate is consistent with supersymmetry, because the $F$-term
equation of motion $<\Phi \bar{\Phi}> - m=0$
is satisfied. Notice that this preservation of symmetry when $m\not=0$ is in agreement with the
kappa symmetry of our brane probes.

Apart from all this, there is a vectorial $U(1)$ symmetry that remains 
unbroken in our brane probe analysis and that we have associated with the invariance under
translations in  $\tilde\varphi$. On the field theory side this symmetry can be identified
with a phase change of the full chiral/antichiral multiplet 
$\Phi \to e^{i\alpha} \Phi,\,\,\bar\Phi \to e^{-i   \alpha} \bar\Phi$, which is nothing
but the  $U(1)_B$ baryonic number symmetry of the theory. The two possible assignations
$\pm 1$ of the baryonic charge are in correspondence with the two possible
identifications of the $S^1$ described by $\tilde\varphi$ with the $S^1$
parametrized by $\varphi$. The spectrum we have found is independent of this
identification.

We would like also to comment briefly on the possibility of taking the 
parameter $r_*=0$. Given that this parameter can be associated with the 
mass of the quark (since it is the characteristic distance between the 
probe brane and the background) one would like to study the case in which 
this parameter is taken to be zero. Nevertheless the approach implemented 
here seems to break down for this particular value of the parameter. 
Indeed, taking $r_*=0$ will imply that $\theta_0=0$, so, a fluctuation of 
it can lead to negative values of $\theta$ taking us out of the range of 
this coordinate. The fact that our approach apparently does not
work for the case of massless quarks seems to be in agreement with the 
fact that SQCD with massless flavors, 
has some special properties like spontaneous breaking of  supersymmetry,
the existence of a runaway potential (Affleck-Dine-Seiberg potential) and the
non-existence of a  well-defined vacuum state. Notice that, when $r_*\not= 0$, 
our approach, by 
construction, deals with massive quarks that preserve supersymmetry, since 
our probes were constructed by that requirement.

\medskip
\section{Summary and conclusions}
\medskip

Let us summarize and repeat here some points that we believe are worth 
stressing again.

We studied a supergravity dual to ${\cal N}=1$ SYM, based on a geometry
 representing a stack of $D5$ branes wrapping a two-cycle. In this gravity 
solution, we have found special surfaces where one can put probe 
D5-branes without spoiling supersymmetry. This was done, basically, by 
imposing that the Killing spinors of the background satisfy the projection
imposed by the kappa symmetry of the D5-brane probe.

A wide variety of kappa symmetric solutions was found and a rich 
mathematical structure was pointed out. In fact, solutions where the 
probes are at a fixed distance from the "background branes" are shown to 
break supersymmetry. This phenomenon is in agreement with the non-existence 
of moduli space of  ${\cal N}=1$ SYM. On the other hand, we studied 
solutions corresponding to the abelian background (that is to say the 
large $r$ regime of the full background). These abelian solutions are 
shown to have a very interesting analytic structure
(harmonic functions and Cauchy-Riemann equations show up) allowing interesting 
explicit solutions. In extending these studies to the full background, we 
found several classes of non-abelian solitons. We think that it might be possible that
these solutions show  themselves useful when studying other aspects of the model not
explored in  this paper.

Then, in section 7, we have used the kappa symmetric solutions (in 
particular those that we called unit-winding) to introduce fundamental 
matter in the dual to SYM theory.

Given that the brane probes do not backreact on the background, these 
flavors are introduced in the so called quenched approximation. 
Nevertheless, many qualitative features and quantitative predictions 
for the strong coupling regime of  ${\cal N}=1$ SQCD can 
be addressed. Among them, the qualitative difference between the massless and 
massive flavors are clear in this picture. Indeed, by construction, our 
approach deals with the massive-flavor case, so, the problems or 
peculiarities of the massless case can be seen in this approach under a 
different geometrical perspective.

Other characteristic feature of SQCD with few flavors is the breaking 
pattern of R-symmetry. The $\ZZ_{2N} \to \ZZ_{2}$ breaking  is
geometrically very  clear from the brane probe perspective. Also, the preserved $U(1)_B$
is  geometrically realized, as explained in the previous section, by arbitrary 
changes in the coordinate $\tilde{\varphi}$.

A mass spectrum for the low energy excitations (mesons) of massive SQCD
was found and we believe this is a very interesting and quantitative 
prediction of this 
paper.
In fact, a nice formula for the masses is derived that exhibits a BPS-like 
behavior with the level ($n$) and  with the Kaluza-Klein quantum number ($l$) of the
meson.  Basically, our formula gives the meson masses in terms of
the mass of the fundamentals and the Casimir energy due to finite size 
of the meson,  and shows explicitly the contamination of the meson spectrum due to the 
Kaluza-Klein modes. It would be very 
interesting if lattice calculations could validate or invalidate the 
formula  found here.

The appendices provide  a hopefully very useful technical information  
for other workers in the area  which contain, in particular, 
the explicit expressions of the Killing spinors of the background considered here
and for another related background dual to  ${\cal N}=1$ SYM theory with a Chern-Simons 
term in $2+1$ dimensions.

We believe that this paper opens some interesting routes of future 
research. To mention some of them, it should be interesting to repeat the 
procedure here in other ``wrapped branes" setups, dual to lower 
supersymmetric field theories, to see if the 
features for the mass 
formula and symmetries are  preserved. In particular, it would be very interesting to
apply the methods employed here to the Klebanov-Strassler background \cite{KS}. The
analysis of ref. \cite{KSglueball} of the glueball spectrum suggests that, contrary to
what happens to the background studied here,  the spectrum of
mesons could be obtained without introducing any additional cut-off. Actually, this meson
spectrum was obtained in ref. \cite{Sonnen} from the fluctuations of a D7-brane probe
wrapped on a submanifold which is  analogous to our cylinders. Presumably, our method
would allow us to obtain a family of supersymmetric embeddings which depend on a parameter
that can be identified with the distance between the brane probe and the origin.

On the other hand, the brane probe lagrangian  should be regarded  as a sort of ``low 
energy 
effective lagrangian" for the strong coupling limit of SQCD with few 
flavors; it would be very nice to develop the physics of this effective 
theory because,  possibly, the scattering of mesons
and other interesting features, like low energy theorems,  could  be 
studied or understood in this geometrical context.

Extending the treatment of this paper to non supersymmetric theories might 
seem a contradiction (the paper is based on kappa symmetry)
but it is likely that with softly broken backgrounds like those
appeared recently in the literature, see ref. \cite{SeveralMN}, similar methods to the
ones exhibited  here can be used.

\medskip
\section*{Acknowledgments}
\medskip
We are grateful to D. Crooks, J. D. Edelstein, J. Erdmenger, S. A. Hartnoll, R.
Hern\'andez, M. Kruczenski~, J. M. Maldacena, D. \& T. Mateos, J. Mas, J. W. Negele, L.
Pando-Zayas, R. Portugues, K. Rajagopal, M. Schvellinger and D. Tong for useful
discussions and encouragement. We have benefited of the nice atmosphere of the Benasque
Center for Science, where part of this work has been done.  C. N. is supported by a
Pappalardo Fellowship, and in part by funds  provided by the U.S. Department of Energy
(D.O.E) under cooperative  research agreement \# DF-FC02-94ER40818. A. P. and A. V. R are
supported in part by MCyT,
FEDER and Xunta de Galicia under grant  BFM2002-03881 and by  the EC Commission under the
FP5 grant HPRN-CT-2002-00325.

\vskip 1cm
\renewcommand{\theequation}{\rm{A}.\arabic{equation}}
\setcounter{equation}{0}
\medskip
\appendix
\section{Supersymmetry of the ${\cal N}=1$ supergravity duals}
\medskip
The supersymmetry transformations for the type IIB dilatino $\lambda$  and 
gravitino $\Psi_{\mu}$ when the RR three-form is nonzero are:
\bear
\delta\lambda&=&{i\over
2}\,\,\partial_{\mu}\,\phi\,\Gamma^{\mu}\,\epsilon^*\,+\, {1\over
24}\,e^{{\phi\over 2}}\,
F^{(3)}_{\mu_1\mu_2\mu_3}\,\Gamma^{\mu_1\mu_2\mu_3}\,\epsilon\,\,,\rc\rc
\delta\Psi_{\mu}&=&D_{\mu}\epsilon\,+\,{i\over 96}\,
e^{{\phi\over 2}}\,F^{(3)}_{\mu_1\mu_2\mu_3}\,
\Big(\,\Gamma_{\mu}^{\,\,\mu_1\mu_2\mu_3}\,-9\,\delta_{\mu}^{\mu_1}\,
\Gamma^{\mu_2\mu_3}\,\Big)\,\epsilon^*\,\,.
\label{SUSYIIB}
\eear
The Killing spinors of the background are those $\epsilon$ for which the right-hand side
of eq.  (\ref{SUSYIIB}) vanishes. In order to satisfy the equations 
$\delta\lambda=\delta\Psi_{\mu}=0$ we will have to impose certain projection conditions
to $\epsilon$. Once these projections are imposed, the equations 
$\delta\lambda=\delta\Psi_{\mu}=0$ are equivalent to a system of first-order
differential equations for the metric and RR three-form. Moreover, following the
methodology of ref. \cite{Twist} (see also ref. \cite{Apreda}), we will be able to
determine the explicit form of $\epsilon$ from the projections that satisfies. In this
appendix we will carry out this program for two ${\cal N}=1$ supergravity duals which
correspond to D5-branes wrapping a two- and a three-cycle.

\subsection{Fivebranes wrapped on a two-cycle}
\medskip
For a metric ansatz such as that of eq. (\ref{metric}), 
let us consider the frame
\bear
e^{x^i}&=&e^{{\phi\over 4}}\,d x^i\,\,,
\,\,\,\,\,\,\,(i=0,1,2,3)\,\,,\rc\rc
e^{1}&=&e^{{\phi\over 4}+h}\,d\theta\,\,,
\,\,\,\,\,\,\,\,\,\,\,\,\,\,
e^{2}=e^{{\phi\over 4}+h}\,\sin\theta d\varphi\,\,,\rc\rc
e^{r}&=&e^{{\phi\over 4}}\,dr\,\,,
\,\,\,\,\,\,\,\,\,\,\,\,\,\,
e^{\hat i}={e^{{\phi\over 4}}\over 2}\,\,
(\,w^i\,-\,A^i\,)\,\,,\,\,\,\,\,\,\,(i=1,2,3)\,\,.
\label{frame}
\eear
In order to solve the equations $\delta\lambda=\delta\Psi_{\mu}=0$, 
we impose to the spinor $\epsilon$ the condition for a SUSY two-cycle:
\beq
\Gamma_{12}\,\epsilon\,=\,\hat\Gamma_{12}\,\epsilon\,\,,
\label{projectionone}
\eeq
and the following projection
\beq
\epsilon\,=\,i\epsilon^*\,\,.
\label{projectiontwo}
\eeq
Then, the condition $\delta\lambda=0$ yields the equation
\beq
\phi'\,\epsilon\,-\,\big(\,1\,+\,{e^{-2h}\over 4}\,(a^2-1)\,\big)\,
\Gamma_r\,\hat\Gamma_{123}\,\epsilon\,+\,{1\over 2}\,a'\,e^{-h}\,
\Gamma_2\hat \Gamma_2\epsilon\,=\,0\,\,.
\label{dilatino}
\eeq
Let us now consider the supersymmetric variation of the different components of
the gravitino. It is easy to verify that, with the projections
(\ref{projectionone}) and (\ref{projectiontwo}), the equation corresponding
to the unwrapped directions $x^i$ and that corresponding to the directions
$\hat i$,  coincide with the one obtained from the variation of the dilatino
(eq. (\ref{dilatino})). On the other hand the variation of the components along
the two-sphere gives rise to the following equation
\beq
h'\epsilon\,-\,{1\over 2}\,a'\,e^{-h}\,\Gamma_2\hat\Gamma_2\epsilon\,-\,
a\,e^{-h}\,\Gamma_2\hat\Gamma_2\Gamma_r\hat\Gamma_{123}\,\epsilon\,+\,
{1\over 2}\,(a^2-1)\,e^{-2h}\,\Gamma_r\hat\Gamma_{123}\,\epsilon\,=\,0\,\,,
\label{gravitino}
\eeq
whereas, after using eq. (\ref{dilatino}),  the condition $\delta\psi_r=0$ is
equivalent to
\beq
\partial_r\epsilon\,-\,{1\over 2}\,a'\,e^{-h}\,\Gamma_2\hat\Gamma_2\,
\epsilon\,-\,{1\over 8}\phi'\,\epsilon\,=\,0\,\,.
\label{psir}
\eeq
Moreover, from the variation of the dilatino (eq. (\ref{dilatino})) we learn that 
$\epsilon$ must satisfy
\beq
\Gamma_{r}\hat\Gamma_{123}\,\epsilon\,=\,
\big(\,\beta+\tilde\beta\Gamma_2\hat\Gamma_2
\,\big)\,\epsilon\,\,,
\label{radial}
\eeq
where $\beta$ and $\tilde\beta$ can be read from eq. (\ref{dilatino}), namely
\beq
\beta\,=\,{\phi'\over 1+{e^{-2h}\over 4}\,(a^2-1)}\,\,,
\,\,\,\,\,\,\,\,\,\,\,
\tilde\beta\,=\,{1\over 2}\,
{e^{-h}a'\over 1+{e^{-2h}\over 4}\,(a^2-1)}\,\,.
\label{beta}
\eeq
On the other hand, from the condition 
$(\Gamma_{r}\hat\Gamma_{123})^2\epsilon=\epsilon$ and since 
$\{\Gamma_{r}\hat\Gamma_{123},\Gamma_2\hat\Gamma_2\}=0$, it follows that:
\beq
\beta^2+\tilde\beta^2=1\,\,,
\eeq
and, therefore, we can represent $\beta$ and $\tilde\beta$ as:
\beq
\beta=\cos\alpha\,,\,\,\,\,\,\,\,\,\,\,\,
\tilde\beta=\sin\alpha\,\,.
\eeq
Substituting the radial projection (eq. (\ref{radial})) into eq.
(\ref{gravitino}) and considering the terms with and without 
$\Gamma_2\hat\Gamma_2$, we get the following two equations:
\beq
h'\,=\,-{1\over 2}\,e^{-2h}\,(a^2-1)\,\beta-ae^{-h}\tilde\beta\,\,,
\,\,\,\,\,\,\,\,\,\,\,
a'\,=\,-2a\beta\,+\,e^{-h}\,(a^2-1)\,\tilde\beta\,\,.
\label{gravitinodos}
\eeq
By using the definition of $\beta$  and $\tilde\beta$ (eq. (\ref{beta})) into the second
equation  in (\ref{gravitinodos}), we get the following relation between $\phi'$  and $a'$:
\beq
\phi'\,=\,-{a'\over 2a}\,\Big[\,1\,-\,{1\over 4}\,e^{-2h}\,
(a^2-1)\,\Big]\,\,.
\label{phia}
\eeq
Furthermore, 
from the condition $\beta^2+\tilde\beta^2=1$ one obtains a new relation between
$\phi'$ and $a'$, namely:
\beq
\phi'^2\,+\,{1\over 4}\,e^{-2h}\,a'^2\,=\,
[\,1\,+\,{1\over 4}\,e^{-2h}\,(\,a^2-1\,)\,]^2\,\,.
\label{square}
\eeq
By combining eqs. (\ref{phia}) and  (\ref{square}) one can get 
the expression of $\phi'$ and $a'$ in terms of $a$ and $h$. 
Moreover, by using these results in eq. (\ref{beta}),
one can get
$\beta$ and $\tilde\beta$ as functions of $a$ and $h$ and, by plugging the
corresponding expressions on the first eq. in (\ref{gravitinodos}), one can
obtain the differential equation for $h$. In order to write these expressions
in a compact form, let us define:
\beq
Q\,\equiv\,\sqrt{e^{4h}\,+\,{1\over 2}\,e^{2h}\,(a^2+1)\,+\,{1\over 16}\,
(a^2-1)^2}\,\,.
\label{defQ}
\eeq
Then, one has the following system of first-order differential equations for
$\phi$, $h$ and $a$:
\bear
\phi'&=&{1\over Q}\,
\Big[\,e^{2h}\,-\,{e^{- 2h}\over 16}\,(a^2-1)^2\,\Big]\,\,,\rc\rc
h'&=&{1\over 2Q}\,
\Big[\,a^2+1+{e^{- 2h}\over 4}\,(a^2-1)^2\,\Big]\,\,,\rc\rc
a'&=&-{2a\over Q}\,
\Big[\,e^{ 2h}+{1\over 4}\,(a^2-1)\,\Big]\,\,,
\label{differentialeqs}
\eear 
and the values of $\beta=\cos\alpha$ and $\tilde\beta=\sin\alpha$, which are given by:
\beq
\sin\alpha\,=\,-{ae^h\over Q}\,\,,
\,\,\,\,\,\,\,\,\,\,\,\,
\cos\alpha\,=\,{e^{2h}\,-\,{1\over 4}\,(\,a^2-1\,)\over Q}\,\,.
\label{alpha}
\eeq
It is interesting to notice that, when solving the quadratic eq. (\ref{square})
to obtain (\ref{differentialeqs}) and (\ref{alpha}), we have a sign ambiguity.
We have fixed this sign by requiring that $h'$ is always positive.  

It remains to verify the fulfillment of equation (\ref{psir}). Notice first of all
that the radial projection (\ref{radial}) can be written as
\beq
\Gamma_r\,\hat\Gamma_{123}\,\epsilon\,=\,e^{\alpha\Gamma_2\hat\Gamma_2}\,
\epsilon\,\,,
\eeq
which, after taking into account that 
$\{\Gamma_{r}\hat\Gamma_{123},\Gamma_2\hat\Gamma_2\}=0$,  can be solved as:
\beq
\epsilon\,=\,e^{-{\alpha\over 2}\,\Gamma_2\hat\Gamma_2}\,\,\epsilon_0\,\,,
\,\,\,\,\,\,\,\,\,\,\,\,\,\,\,\,\,\,\,\,
\Gamma_r\,\hat\Gamma_{123}\,\epsilon_0=\epsilon_0\,\,.
\eeq
Inserting this parametrization of $\epsilon$ into eq. (\ref{psir}), we get two
types of terms, with and without $\Gamma_2\hat\Gamma_2$, which yield the
following two equations:
\beq
\partial_r\epsilon_0-{1\over 8}\,\phi'\,\epsilon_0\,=\,0\,\,,
\,\,\,\,\,\,\,\,\,\,\,\,\,\,\,\,\,\,\,\,
\alpha'\,=\,-e^{-h}\,a'\,\,.
\label{phase}
\eeq
The first of these equations can be solved immediately, namely
\beq
\epsilon_0\,=\,e^{{\phi\over 8}}\,\eta\,\,,
\eeq
with $\eta$ being a constant spinor. Moreover, by differentiating eq.
(\ref{alpha}) and using eq. (\ref{differentialeqs}), one can verify
that the second equation in (\ref{phase}) is satisfied. 
Thus, the explicit form of the Killing spinor is
\beq
\epsilon\,=\,e^{{\alpha\over 2}\,\Gamma_1\hat\Gamma_1}\,\,
e^{{\phi\over 8}}\,\,\eta\,\,,
\eeq
where $\eta$ is a constant spinor satisfying:
\beq
\Gamma_{x^0\cdots x^3}\,\Gamma_{12}\,\eta\,=\,\eta\,\,,
\,\,\,\,\,\,\,\,\,\,\,\,
\Gamma_{12}\,\eta\,=\,\hat\Gamma_{12}\,\eta\,\,,
\,\,\,\,\,\,\,\,\,\,\,\,
\eta\,=\,i\eta^*\,\,.
\eeq
Eq. (\ref{differentialeqs}) is a system of first-order differential equations whose
solution determines the metric, dilaton and RR three-form of the background. It can
be verified that the values given in eq. (\ref{MNsol}) solve the system
(\ref{differentialeqs}). Moreover, by plugging in eq. (\ref{defQ}) the values of $h$ and
$a$ given in eq. (\ref{MNsol}),  one verifies that
\beq
Q=r\,\,.
\eeq
Then, one gets the following simple expression for $\cos\alpha$:
\beq
\cos\alpha\,=\,{\rm \coth} 2r\,-\,{2r\over \sinh^22r}\,\,.
\label{alphaexplicit}
\eeq
On the other hand, being a spinor of definite chirality of type IIB supergravity, 
$\epsilon$ satisfies  $\Gamma_{x^0\cdots
x^3}\Gamma_{12}\Gamma_{r}\hat\Gamma_{123}\epsilon=\epsilon$. Thus, if we multiply the
radial projection condition (\ref{radial}) by  $\Gamma_{x^0\cdots x^3}\Gamma_{12}$, we
obtain:
\beq
\Gamma_{x^0\cdots x^3}\,\big(\,
\cos\alpha\,\Gamma_{12}\,+\,\sin\alpha\,\Gamma_1\hat\Gamma_2\,)\,
\epsilon\,=\,\epsilon\,\,.
\label{alphaproj}
\eeq
Eqs. (\ref{projectionone}),  (\ref{projectiontwo}) and (\ref{alphaproj}) are the ones
needed in the analysis of the kappa symmetry of a D5-brane probe (eq.
(\ref{fullprojection})). 

\subsection{Fivebranes wrapped on a three-cycle}
\medskip

As a second example of the calculation of the Killing spinors of a ${\cal N}=1$
supergravity dual, let us consider the supergravity solution for a D5 brane wrapped
on a three-sphere. In this case the ansatz for the metric in Einstein frame
is
\beq
ds^2\,=\,e^{{\phi\over 2}}\,
\big[\,dx_{1,2}^{2}\,+\,{1\over 4}\,R(r)^2\,(\,\sigma^i\,)^2\,+\,
dr^2\,+\,{1\over 4}\,(\,w^i\,-\,A^i\,)^2\,\big]\,\,,
\eeq
where $\phi$ is the dilaton, $R(r)$ is a function to be determined and
$\sigma^i$ and $w^i$ ($i=1,2,3$) are two sets of $su(2)$
left invariant one-forms. The gauge field components
$A^i$ are parametrized  in terms of a new function $\omega(r)$ as:
\beq
A^i\,=\,{1+\omega(r)\over 2}\,\,\sigma^i\,\,.
\eeq
Let $F^i$ be the field strength of $A^i$ (defined as in eq. (\ref{fieldstrenght})). Its
components are:
\beq
F^i\,=\,{\omega'\over 2}\,dr\wedge \sigma^i\,+\,
{\omega^2-1\over 8}\,\epsilon_{ijk}\,\sigma^j\wedge\sigma^k\,\,.
\eeq
The RR three-form of the background is now
\beq
F_{(3)}\,=\,-{1\over 4}\,
(w^1-A^1)\wedge (w^2-A^2) \wedge (w^3-A^3)\,+\,
{1\over 4}\,\sum_i\,F^i \wedge (w^i-A^i)\,+\,h\,\,,
\eeq
where $h$ is determined by requiring that $F_{(3)}$ satisfies
the Bianchi identity $dF_{(3)}=0$. One easily verifies that $h$
must satisfy the following equation
\beq
dh\,=\,{1\over 4}\,\sum_i\,F^i\wedge F^i\,\,.
\eeq
By explicit calculation, one gets:
\beq
\sum_i\,F^i\wedge F^i\,=\,{1\over 8}\,
(\,\omega^2\,-\,1\,)\,\omega'\,\,
\epsilon_{jkm}\,dr\wedge\sigma^j\wedge\sigma^k\wedge\sigma^m\,\,.
\eeq
Thus, the equation for $h$ can be solved as:
\beq
h\,=\,{1\over 32}\,{1\over 3!}\,V(r)\,
\epsilon_{ijk}\,\sigma^i\wedge\sigma^j\wedge\sigma^k\,\,,
\eeq
where
\beq
V(r)\,=\,2\omega^3\,-\,6\omega\,+\,8k\,\,,
\eeq
with $k$ being a constant. 

Let us study the supersymmetry preserved by this ansatz in the
frame:
\bear
&&e^{x^i}\,=\,e^{{\phi\over 4}}\,dx^i\,\,,
\,\,\,\,\,\,\,\,\,\,\,\,\,\,
(i=0,1,2)\,\,,
\,\,\,\,\,\,\,\,\,\,\,\,\,\,\,\,\,\,\,\,\,\,\,\,\,\,\,\,
e^r\,=\,e^{{\phi\over 4}}\,dr\,\,,\rc\rc
&&e^{i}\,=\,{1\over 2}\,\,e^{{\phi\over 4}}\,\,R
\,\sigma^i\,\,,
\,\,\,\,\,\,\,\,\,\,\,\,\,\,\,\,\,\,\,\,\,\,\,\,\,\,\,\,
\,\,\,\,\,\,\,\,\,\,\,\,\,\,\,\,\,\,\,\,\,\,\,\,\,\,\,\,
e^{\hat i}\,=\,{1\over 2}\,\,e^{{\phi\over 4}}\,\,
(w^i-A^i)\,\,,
\,\,\,\,\,\,\,\,\,\,\,\,\,\,
(i=1,2,3)\,\,,\rc\rc
\eear
We now impose the following projections on the spinors
\bear
&&\Gamma_1\hat\Gamma_1\,\epsilon\,=\,
\Gamma_2\hat\Gamma_2\,\epsilon\,=\,
\Gamma_3\hat\Gamma_3\,\epsilon\,\,,\rc\rc
&&\epsilon\,=\,i\epsilon^*\,\,.
\eear
Then, the condition $\delta\lambda\,=\,0$ becomes
\beq
\phi'\,\epsilon\,-\,\bigg(\,1\,+\,3\,{\omega^2\,-\,1\over 4R^2}
\bigg)\,\Gamma_r\,\hat\Gamma_{123}\,\epsilon\,+\,
{3\omega'\over 4R}\,\Gamma_1\hat\Gamma_1\epsilon\,-\,
{V\over 8R^3}\,\Gamma_1\hat\Gamma_1\,
\Gamma_r\,\hat\Gamma_{123}\,\epsilon\,=\,0\,\,.
\label{MNasdilatino}
\eeq
Moreover,
from $\delta\psi_i=0$, one gets
\beq
R'\epsilon\,-\,{\omega'\over 2}\,
\Gamma_1\hat\Gamma_1\epsilon\,+\,
{\omega^2-1\over R}\,\Gamma_r\,\hat\Gamma_{123}\,\epsilon\,+\,
\Big(\,{V\over 8R^2}\,-\,\omega\,\Big)\,
\Gamma_1\hat\Gamma_1\,
\Gamma_r\,\hat\Gamma_{123}\,\epsilon\,=\,0\,\,.
\label{MNasgravitino}
\eeq
The vanishing of the supersymmetric variation of the radial component
of the gravitino gives rise to:
\beq
\partial_r\epsilon\,-\,{3\omega'\over 4R}\,
\Gamma_1\hat\Gamma_1\epsilon\,-\,{1\over 8}\,\phi'\,\epsilon\,=\,0\,\,,
\label{MNasgravir}
\eeq
where we have used eq. (\ref{MNasdilatino}). 
Let us solve these equations by taking the projection
\beq
\Gamma_r\,\hat\Gamma_{123}\,\epsilon\,=\,\big(\,
\beta\,+\,\tilde\beta\Gamma_1\hat\Gamma_1\,\big)\,\epsilon\,\,.
\label{MNasrad}
\eeq
As in case studied in section (A.1), from 
$(\Gamma_r\,\hat\Gamma_{123})^2\,\epsilon\,=\,\epsilon$ and 
$\{\Gamma_r\,\hat\Gamma_{123}, \Gamma_1\hat\Gamma_1\}=0$ 
it follows  that
$\beta^2+\tilde\beta^2\,=\,1$ and thus we can take $\beta=\cos\alpha$
and $\tilde\beta=\sin\alpha$.

Let us substitute our ansatz for
$\Gamma_r\,\hat\Gamma_{123}\,\epsilon$ (eq. (\ref{MNasrad}))
on the equations coming from
the dilatino and gravitino (eqs. (\ref{MNasdilatino}) and (\ref{MNasgravitino})). From the
terms containing the unit matrix, we obtain equations for $\phi'$ and $R'$:
\bear
\phi'&=&\Big(\,1\,+\,3\,{\omega^2-1\over 4R^2}\,\Big)\,\beta\,-\,
{V\over 8R^3}\,\tilde\beta\,\,,\rc\rc
R'&=&{1-\omega^2\over R}\,\beta\,+\,\Big(\,{V\over 8R^2}\,-\,\omega
\,\Big)\,\tilde\beta\,\,.
\eear
Moreover, by considering the terms with $\Gamma_1\hat\Gamma_1$, we
obtain two expressions for $\omega'$
\bear
3\omega'&=&{V\over 2R^2}\,\beta\,+\,\Big(\,4R\,+\,3\,
{\omega^2-1\over R}\,\Big)\,\tilde\beta\,\,,\rc\rc
\omega'&=&\Big(\,{V\over 4R^2}\,-\,2\omega\,\Big)\,\beta\,+\,
2\,{\omega^2-1\over R}\,\tilde\beta\,\,.
\eear
By combining these last two equations we get
\beq
\Big(\,{V\over 24R^2}\,-\,w\,\Big)\,\beta\,=\,
\Big(\,{1-w^2\over 2R}\,+\,{2R\over 3}\,\Big)\,\tilde\beta\,\,.
\eeq
By plugging this last relation in the condition 
$\beta^2+\tilde\beta^2\,=\,1$ one can easily obtain the expression of
$\beta$ and $\tilde\beta$. Indeed, let us define $M$  as follows:
\beq
M\,\equiv\,\Big(\,{V\over 24R^2}\,-\,w\,\Big)^2\,+\,
\Big(\,{1-w^2\over 2R}\,+\,{2R\over 3}\,\Big)^2\,\,.
\eeq
In terms of this new quantity $M$, the coefficients $\beta$ and
$\tilde\beta$ are given by:
\beq
\beta\,=\,\cos\alpha\,=\,{1\over\sqrt{M}}\,\,
\Bigg(\,
{2R\over 3}\,+\,{1-\omega^2\over 2R}\,\Bigg)
\,\,,
\,\,\,\,\,\,\,\,\,\,\,\,\,\,\,\,\,\,
\tilde\beta\,=\,\sin\alpha\,=\,{1\over\sqrt{M}}\,\,
\Bigg(\,
{V\over 24R^2}\,-\,\omega
\,\Bigg)\,\,.
\eeq
By using these values of $\beta$ and
$\tilde\beta$ in the equations which determine $R\,'$, $\omega\,'$ and
$\phi'$, we obtain a system of first-order
BPS equations which are identical  to those written in refs. \cite{Martin,MaldaNas,CVdos}
 (see also ref. \cite{JaumeG}). They are:
\bear
R\,'&=&{1\over 3\sqrt{M}}\,\Big[\,{V^2\over 64R^4}\,+\,
{1\over
2R^2}\,\Big(\,3(1-\omega^2)^2\,-\,V\omega\,\Big)
\,+\,\omega^2\,+\,2\,\Big]\,\,,\rc\rc
\omega\,'&=&{4R\over 3\sqrt{M}}\,\Big[\,{V\over
32R^4}\,(\,1-\omega^2\,)\,+\, {2k\,-\,\omega^3\over
2R^2}\,-\,\omega\,\Big]\,\,,\rc\rc
\phi'&=&-{3\over 2}\,\big(\,\log R\,\big)'\,+\,{3\over 2}\,
{\sqrt{M}\over R}\,\,.
\label{MNasDE}
\eear
The radial projection condition (\ref{MNasrad}) can be written as
\beq
\Gamma_r\,\hat\Gamma_{123}\,\epsilon\,=\,
e^{\alpha\Gamma_1\hat\Gamma_1}\,\epsilon\,\,.
\eeq
Since $\{\Gamma_r\,\hat\Gamma_{123}, \Gamma_1\hat\Gamma_1\}=0$, this
equation can be solved as: 
\beq
\epsilon\,=\,
e^{-{\alpha\over 2}\,\Gamma_1\hat\Gamma_1}\,\epsilon_0\,\,,
\,\,\,\,\,\,\,\,\,\,\,\,\,\,\,
\Gamma_r\,\hat\Gamma_{123}\,\epsilon_0\,=\,\epsilon_0\,\,.
\eeq
Plugging now this parametrization of $\epsilon$ into the equation
obtained from the variation of the radial component of the gravitino
(eq. (\ref{MNasgravir}) ), we arrive at the  following two equations:
\beq
\partial_r\,\epsilon_0\,=\,{\phi'\over 8}\,\epsilon_0\,\,,
\,\,\,\,\,\,\,\,\,\,\,\,\,\,\,\,\,\,
\alpha'\,=\,-{3\over 2}\,\omega'\,R\,\,.
\eeq
The equation for $\epsilon_0$ can be solved immediately:
\beq
\epsilon_0\,=\,e^{{\phi\over 8}}\,\eta\,\,,
\eeq
where $\eta$ is a constant spinor. Moreover, one can verify that 
the equation for $\alpha$ is a consequence of the first-order BPS
equations (\ref{MNasDE}). Therefore, the Killing spinors for this geometry are of the
form:
\beq
\epsilon\,=\,e^{-{\alpha\over 2}\,\Gamma_1\hat\Gamma_1}\,\,
e^{{\phi\over 8}}\,\eta\,\,,
\eeq
where $\eta$ is constant and satisfies the following conditions:
\bear
&&\Gamma_{x^0\cdots x^2}\,\Gamma_{123}\,\eta\,=\,\eta\,\,,\rc\rc
&&\Gamma_1\hat\Gamma_1\,\eta\,=\,\Gamma_2\hat\Gamma_2\,\eta\,=\,
\Gamma_3\hat\Gamma_3\,\eta\,\,,\rc\rc
&&\eta\,=\,i\eta^*\,\,.
\eear
Notice that, for the spinor $\epsilon$,  the first of these projections
can be recast in the form:
\beq
\Gamma_{x^0\cdots x^2}\,\Big(\,\cos\,\alpha\Gamma_{123}\,-\,
\sin\alpha\hat\Gamma_{123}\,\Big)\,\epsilon\,=\,\epsilon\,\,.
\eeq

\renewcommand{\theequation}{\rm{B}.\arabic{equation}}
\setcounter{equation}{0}
\section{Asymptotic behaviour of the fluctuations}
\medskip
This appendix is devoted to the determination of the asymptotic form of the fluctuations
around the kappa-symmetric static configurations of the D5-brane probe. In subsection B.1
we will consider the large and small $r$ behaviour of the solutions of the equation of
motion corresponding to the lagrangian (\ref{lagrafluct}), which describes the small
oscillations around the unit-winding embedding (\ref{thetazero}). 
In section B.2 we will study the asymptotic form of the fluctuations of the
$n$-winding embedding in the abelian background. Following our general arguments, the UV
behaviour of the abelian and non-abelian fluctuations must be the same and, thus, 
from this analysis we can get an idea of the nature of the small oscillations around
the general non-abelian $n$-winding configurations (whose analytical form we have not
determined) for large values of the radial coordinate.

\subsection{Non abelian unit-winding embedding}

For large $r$ the  lagrangian (\ref{lagrafluct}) takes the form:
\beq
{\cal L}\,=\,-{1\over 2}\,e^{2\phi}\,\big[\,
r\,(\partial_r\chi)^2\,+\,2r\,\chi\,\partial_r\chi\,+\,r\chi^2\,+\,r\,
(\partial_x\chi)^2\,+\,r\,(\partial_{\varphi}\chi)^2\,\big]\,\,,
\eeq
where we have not expanded the dilaton and we have eliminated the exponentially
suppressed terms. Using the asymptotic value of $\partial_r\phi$:
\beq
\partial_r\phi\,\approx\,1\,-{1\over 4r-1}\,\,,
\eeq
we obtain the following equation for $\xi$:
\beq
\partial^2_r\,\xi\,+\,{8r^2-1\over 4r^2-r}\,
\partial_r\,\,\xi\,+\,
\Bigg(M^2-l^2+1\,+\,{2r-1\over 4r^2-r}
\Bigg)\,\,\xi\,=\,0\,\,.
\eeq
To study the UV behaviour of the solutions of this differential equation it is interesting
to rewrite it with the different coefficient functions expanded in powers of 
$1/r$ as:
\beq
\partial^2_r\,\xi\,+\,\big(\,a_0+
{a_1\over r}+{a_2\over r^2}\,+\,\cdots\big)\,\partial_r\,\xi\,+\,
\big(\,b_0+{b_1\over r}+{b_2\over r^2}\,+\,\cdots\big)\,\xi\,=\,0\,\,,
\label{UVexpand}
\eeq
where
\beq
a_0\,=\,2\,,\,\,\,\,\,\,\,\,
b_0\,=\,M^2-l^2+1\,,\,\,\,\,\,\,\,\,\,\,\,\,\,\,\,\,\,\,\,\,
a_1\,=\,b_1\,=\,{1\over 2}\,,\,\,\,\,\,\,\,\,\,\,\,\,\,\,\,
a_2\,=\,b_2\,=\,-{1\over 8}\,,\cdots\rc\rc
\eeq
We want to solve  eq. (\ref{UVexpand}) by means of an asymptotic Frobenius expansion
of the type 
$\xi=r^{\rho}\,(c_0+c_1/r+\cdots)$ for some exponent $\rho$. By substituting
this expansion on the right-hand side of eq. (\ref{UVexpand}) and comparing the terms
with the different powers of $r$, we notice that there is only one term with
$r^{\rho}$, namely $b_0c_0r^{\rho}$, which cannot be canceled. In order to get
rid of this term, let us define a new function $w$ as
\beq
\xi\,=\,e^{\alpha r}\,w\,\,,
\eeq
with $\alpha$ being a number to be determined. The equation satisfied by $w$ is
the same as that of $\xi$ with the changes:
\bear
&&a_0\rightarrow a_0+2\alpha\,,\,\,\,\,\,\,\,\,\,\,\,\,\,\,\,\,
b_0\rightarrow \alpha^2+\alpha a_0+b_0\,\,,\rc\rc
&&a_i\rightarrow a_i\,\,,\,\,\,\,\,\,\,\,\,\,\,\,\,\,\,\,\,\,\,\,\,\,\,\,\,\,\,
b_i\rightarrow b_i+\alpha a_i\,\,,\,\,\,\,\,\,\,\,
(i=1,2,\cdots)\,\,.
\eear
It is clear that we must impose the condition:
\beq
\alpha^2+\alpha a_0+b_0\,=\,0\,\,,
\label{UVindicial}
\eeq
which  determines the values of  $\alpha$. Writing now 
$w=r^{\rho}\,(c_0+c_1/r+\cdots)$ and looking at the highest power of $r$ in the
equation of $w$ (\ie\ $\rho-1$), we immediately obtain the value of $\rho$,
namely:
\beq
\rho=-{\alpha a_1+b_1\over 2\alpha+a_0}\,\,,
\label{UVfrobenius}
\eeq
and, clearly, as $r\rightarrow\infty$ the asymptotic behaviour of $\xi(r)$ is:
\beq
\xi(r)\approx e^{\alpha r}\,r^{\rho}\,(\,1\,+\,o({1\over r})\,)\,\,.
\eeq
In our case, it is easy to verify that the values of $\alpha$ and $\rho$ are:
\beq
\alpha=-1\pm i\sqrt{M^2-l^2}\,\,,\,\,\,\,\,\,\,\,\,\,\,\,
\rho =-{1\over 4}\,\,.
\eeq
Then, it is clear that we have two independent behaviours for the real function
$\xi(r)$, namely:
\beq
\xi(r)\sim {e^{-r}\over r^{{1\over 4}}}\cos \Big(\,\sqrt{M^2-l^2}\,\,r\,\Big)\,,
\,\,\,\,\,\,
{e^{-r}\over r^{{1\over 4}}}\sin\Big(\,\sqrt{M^2-l^2}\,\,r\,\Big)\,\,.
\label{UVasymp}
\eeq
Notice that all solutions decrease exponentially when $r\rightarrow\infty$. 

Let us now turn to the analysis of the fluctuations for small values of the radial
coordinate. Recall that $r_*\le r\le \infty$. 
Near $r_*$, one can expand $\sin\theta_0$ and $\cos\theta_0$ as follows:
\bear
\sin\theta_0&\approx& 1\,-\,\coth r_*\,(r-r_*)\,+\,{1\over 2}\,
{1+\cosh^2r_*\over \sinh^2r_*}\,(r-r_*)^2\,+\,\cdots\,\,,\rc\rc
\cos\theta_0&\approx&\sqrt{2\coth r_* (r-r_*)}\,\bigg[\,
1\,-{\coth r_*\over 2}\,(1+{1\over 2\cosh^2 r_*})\, (r-r_*)\,+\,\cdots\bigg]\,\,.
\label{sincos}
\eear
Using these expressions it is straightforward to show that 
the lagrangian density of the quadratic fluctuations is given by:
\beq
{\cal L}\,=\,-{1\over 2}\,e^{2\phi_*}\,\,\big[\,
A(r)\,(\,\partial_r\chi)^2\,+\,B(r)2\chi\partial_r\chi\,+\,
C(r)\,\chi^2\,+\,D(r)\,(\,\partial_x\chi)^2\,+\,
E(r)\,(\,\partial_{\varphi}\chi)^2
\,\,\big]\,\,,
\label{IRL}
\eeq
where $\phi_*=\phi(r_*)$ and the functions $A, B, C, D$ and $E$ are of the form:
\bear
A(r)&=&\big[\,8\tanh r_*\,(r-r_*)^3\,\big]^{{1\over 2}}\,\,
\big(\,1\,+\,{\cal A}(r)\,\big)\,\,,\rc\rc
B(r)&=&{\sqrt{2(r-r_*)}\over \sqrt{\coth r_*}}\,\,
\big(\,1\,+\,{\cal B}(r)\,\big)\,\,,\rc\rc
C(r)&=&{1\over \sqrt{2\coth r_*\,(r-r_*)}}\,\,
\big(\,1\,+\,{\cal C}(r)\,\big)\,\,,\rc\rc
D(r)&=&{r_*\over \sqrt{\coth r_*}}\,\sqrt{2(r-r_*)}\,\,
\big(\,1\,+\,{\cal D}(r)\,\big)\,\,,\rc\rc
E(r)&=&\big[\,\tanh r_*\,\big]^{{3\over 2}}\,\,
\sqrt{2(r-r_*)}\,\,\big(\,1\,+\,{\cal E}(r)\,\big)\,\,.
\eear
The functions ${\cal A},{\cal B},{\cal C},{\cal D}, {\cal E}$, satisfy:
\beq
{\cal A}(r),{\cal B}(r), {\cal C}(r),{\cal D}(r)\,{\cal E}(r)\,\,\sim\,\,o(r-r_*)\,\,.
\eeq
Notice that, remarkably, after integrating by parts the 
$\chi\partial_r\chi$ term in the lagrangian, the singular term of $C(r)$
cancels against the leading term of $B(r)$. The equation of motion for
$\xi$ near $r_*$ is:
\beq
\partial^2_r\xi\,+\,{A'(r)\over A(r)}\,\,\partial_r\xi\,+\,
{B'(r)-C(r)+M^2D(r)-l^2E(r)\over A(r)}\,\,\xi\,=\,0\,\,.
\label{IReq}
\eeq
In order to solve this equation in a power series expansion in $r-r_*$, it is
important to understand the singularities of the different coefficients near
$r_*$. It is immediate that:
\beq
{A'(r)\over A(r)}\,=\,{3\over 2}\,\,{1\over r-r_*}\,+\,
{{\cal A}'(r)\over 1+{\cal A}(r)}\,=\, 
{3\over 2}\,\,{1\over r-r_*}\,+\,{\rm regular}\,\,.
\eeq
Similarly, the coefficient of $\xi$ has a simple pole near $r_*$:
\bear
{B'(r)-C(r)+M^2D(r)-l^2E(r)\over A(r)}\,=\,
{3{\cal B}\,'(r_*)-{\cal C}\,'(r_*)+2M^2r_*-2l^2\tanh r_*\over
4}\,{1\over r-r_*}\,+\,{\rm regular}\,\,.\rc
\eear
It follows that the point $r=r_*$ is a singular regular point. The corresponding
Frobenius expansion reads:
\beq
\xi(r)\,=\,(r-r_*)^{\lambda}\,
\sum_{n=0}^{\infty}\,c_n\,(r-r_*)^n\,\,,
\eeq
where $\lambda$ satisfies the indicial equation, which can be obtained by
plugging the expansion in the equation and looking at the term with lowest
power of $r-r_*$ (\ie\ $\lambda-2$). In our case, $\lambda$ must be a root
of the quadratic equation:
\beq
\lambda(\lambda-1)+{3\over 2}\,\lambda\,=\,0\,\,,
\label{IRindicial}
\eeq
\ie:
\beq
\lambda=0, -{1\over 2}\,\,.
\eeq
This means that there are two independent solutions of the differential equation which can
be represented by a Frobenius series around $r=r_*$, one of then is regular as
$r\rightarrow r_*$ (the one corresponding to $\lambda=0$), whereas the  other 
diverges as $(r-r_*)^{-{1\over 2}}$. 

\subsection{ $n$-Winding embeddings in the abelian background}

Let us consider the abelian background and an embedding with winding number
$n$, for which $\partial_{\varphi}\tilde\varphi=n$,
$\sin\theta\partial_{\theta}\tilde\theta\,=\,n\sin\tilde\theta$ and
$\psi=\psi_0={\rm constant}$. Let us define
\bear
V&\equiv&{n\over 2}{\cos\tilde\theta\over \sin\theta}\,+\,{1\over
2}\cot\theta\,=\, {n\over 2\sin\theta}\,\,
{(1+\cos\theta)^n\,-\,(1-\cos\theta)^n\over
(1+\cos\theta)^n\,+\,(1-\cos\theta)^n}\,+\,{1\over 2}\cot\theta\,\,,\rc\rc
W&\equiv&{n\over 2}{\sin\tilde\theta\over \sin\theta}\,=\,
{n(\sin\theta)^{(n-1)}\over 
(1+\cos\theta)^n\,+\,(1-\cos\theta)^n}\,\,.
\eear
Choosing $r$ as worldvolume coordinate and considering
embeddings in which $\theta$ depends
on both $r$ and on the unwrapped coordinates $x$, we obtain the following
lagrangian:
\bear
{\cal L}=-e^{2\phi}\,\sin\theta&\Bigg[
&\sqrt{\big(\,e^{2h}\,+\,V^2\,+\,W^2\,\big)\,\big(\,1\,+\,
(e^{2h}\,+\,W^2)((\partial_r\theta)^2\,+\,(\partial_x\theta)^2)\big)}\,+\cr\cr
&&+\,(e^{2h}\,+\,W^2)\partial_r\theta\,-\,V\,\Bigg]\,\,.
\eear
We shall expand this lagrangian around a configuration $\theta_0(r)$ such that
\beq
e^{2(r-r_*)}\,=\,{1\over 2}\,
{(1+\cos\theta_0(r))^n\,+\,(1-\cos\theta_0(r))^n\over 
(\sin\theta_0(r))^{n+1}}\,\,.
\label{nthetazero}
\eeq
Notice that $\theta_0(r)$ remains unchanged under the transformation
$n\to -n$. Thus, without loss  of generality we shall restrict ourselves 
 from now on to the case $n\ge 0$. Let us now define
\beq
V_0(r)\equiv V|_{\theta(r)=\theta_0(r)}\,\,,
\,\,\,\,\,\,\,\,\,\,\,\,\,\,\,\,
W_0(r)\equiv W|_{\theta(r)=\theta_0(r)}\,\,,
\,\,\,\,\,\,\,\,\,\,\,\,\,\,\,\,
V_{0\theta}(r)\equiv{\partial V\over \partial\theta}
\Big|_{\theta(r)=\theta_0(r)}\,\,.
\eeq
Using 
\beq
\partial_{r}\theta_0\,=\,-{1\over V_0}\,\,,
\eeq
we obtain the following quadratic lagrangian
\bear
{\cal L}\,=\,-{e^{2\phi}\over 2}\,\sin\theta_0\,(e^{2h}\,+\,W_0^2\,)\,
&\Bigg[& {1\over e^{2h}\,+\,V_0^2\,+\,W_0^2}\,
\Big(\,V_0^3\,(\partial_r\chi)^2\,-\,2V_0\,V_{0\theta}\,\chi\partial_r\chi\,+
\rc\rc
&&+\,{V_{0\theta}^2\over V_0}\,\chi^2\,\Big)\,+\,V_0\,(\partial_x\chi)^2
\,\Bigg]\,\,.
\eear
Keeping the leading terms for large $r$, the lagrangian becomes:
\beq
{\cal L}\,=\,-{r\over 2}e^{2\phi}\,\big[\,
{n+1\over 2}\,(\partial_r\chi)^2\,+\,2\chi\partial_r\chi\,+\,
{2\over n+1}\,\chi^2\,+\,{n+1\over 2}\,(\partial_x\chi)^2\,
\big]\,\,,
\eeq
and, if we represent $\chi$ as in eq. (\ref{chiansatz}) with $l=0$, 
the equation of motion for $\xi$ becomes:
\beq
\partial_r^2\xi\,+\,\big(2+{1\over 2r}\big)\,
\partial_r\xi\,+\,\bigg(\,M^2\,+\,{4n\over (n+1)^2}\,+\,{1\over n+1}\,
{1\over r}\,\bigg)\,\xi\,=\,0\,\,.
\eeq
Now we have the following coefficients in eq. (\ref{UVexpand}):
\beq
a_0\,=\,2\,\,,
\,\,\,\,\,\,\,\,\,\,\,\,\,\,\,\,\,\,
b_0\,=\,M^2+{4n\over (n+1)^2}\,\,,
\,\,\,\,\,\,\,\,\,\,\,\,\,\,\,\,\,\,
a_1\,=\,{1\over 2}\,\,,
\,\,\,\,\,\,\,\,\,\,\,\,\,\,\,\,\,\,
b_1\,=\,{1\over n+1}\,\,.
\eeq
By plugging these values in eq. (\ref{UVindicial}) we obtain the following
result for the coefficient $\alpha$ of the exponential:
\beq
\alpha\,=\,-1\,\pm\,
\sqrt{\Big({n-1\over n+1}\Big)^2\,-\,M^2}\,\,.
\eeq
Let us distinguish two cases, depending on the sign inside the square root.
Suppose first that $M^2\ge \big({n-1\over n+1}\big)^2$ and define
\beq
\tilde M^2\,=\,M^2\,-\,\Big({n-1\over n+1}\Big)^2\,\,.
\eeq
In this case the values of the exponent $\rho$ obtained from 
(\ref{UVfrobenius}) are:
\beq
\rho\,=\,-{1\over 4}\,\mp {n-1\over 2(n+1)\tilde M}\,\,i\,\,.
\eeq
It follows that the two real asymptotic solutions are:
\beq
\xi(r)\sim {e^{-r}\over r^{{1\over 4}}}\,
\cos\big[\tilde M r\,-\,{n-1\over 2(n+1)\tilde M}\,\log r\big]\,\,,
\,\,\,\,\,\,\,\,\,\,\,\,
{e^{-r}\over r^{{1\over 4}}}\,
\sin\big[\tilde M r\,-\,{n-1\over 2(n+1)\tilde M}\,\log r\big]\,\,.
\label{nUV}
\eeq
Both solutions vanish exponentially when $r\rightarrow\infty$. 

If $\tilde M^2<0$, let us define $\bar M^2=-\tilde M^2$. In this case $\alpha$
is real, namely $\alpha=-1\pm\bar M$. Notice that $\bar M<1$ and thus
$\alpha<0$. The independent asymptotic solutions are:
\beq
\xi(r)\sim e^{(\bar M-1)r}\,\,
r^{-{1\over 4}\,\big(1\,-\, {n-1\over (n+1)\bar M}\big)}\,\,,
\,\,\,\,\,\,\,\,\,\,\,\,
e^{-(\bar M+1)r}\,\,
r^{-{1\over 4}\,\big(1\,+\, {n-1\over (n+1)\bar M}\big)}\,\,,
\label{nUVotro}
\eeq
and both decrease exponentially, without oscillations, when $r\rightarrow\infty$. This
non-oscillatory character of the functions in eq. (\ref{nUV}) make them inadequate for
the type of boundary conditions we are imposing and, therefore, we shall discard them.

Notice that, for $n=1$, the large $r$ asymptotic solutions (\ref{UVasymp}) and
(\ref{nUV}) coincide. This is of course to be expected since the abelian and non-abelian
configurations coincide in the UV. It is also interesting to compare the magnitude of the
fluctuation with that of the unperturbed configuration for large $r$. By inspecting eq. 
(\ref{nthetazero}) one readily concludes that
\beq
\theta_0(r)\,\sim\,e^{-{2\over n+1}r}\,\,,
\,\,\,\,\,\,\,\,\,
(r\to \infty)\,\,.
\eeq
By comparing this behaviour with eq. (\ref{nUV}) one finds
\beq
{\xi(r)\over \theta_0(r)}\sim {e^{-{n-1\over n+1}r}\over r^{{1\over 4}}}\,\,,
\,\,\,\,\,\,\,\,\,
(r\to \infty)\,\,.
\eeq
Thus, for $n\ge 1$ one has that ${\xi(r)\over \theta_0(r)}\to 0$ as $r\to \infty$. On the
contrary for $n= 0$, both in the abelian and non-abelian case, the ratio 
${\xi(r)\over \theta_0(r)}$ diverges in the UV and the first order expansion breaks down.

Let us now consider the IR behaviour of the fluctuations. 
Near $r_*$ one has to leading order that $\sin\theta_0\approx 1$ and
\bear
&&\cos\theta_0\approx {2\over \sqrt{1+n^2}}\,\sqrt{r-r_*}\,\,,
\,\,\,\,\,\,\,\,\,\,\,\,\,\,\,\,\,\,\,\,\,\,\,\,
V_0\approx \sqrt{n^2+1}\,\sqrt{r-r_*}\,\,,\rc\rc
&&W_0\approx{n\over 2}\,\,,
\,\,\,\,\,\,\,\,\,\,\,\,\,\,\,\,\,\,\,\,\,\,\,\,
\,\,\,\,\,\,\,\,\,\,\,\,\,\,\,\,\,\,\,\,\,\,\,\,
\,\,\,\,\,\,\,\,\,\,\,\,\,\,\,\,\,\,\,\,\,\,\,\,
V_{0\theta}\approx-{n^2+1\over 2}\,\,.
\eear
The IR lagrangian is of the same form as in eq. (\ref{IRL}) (with $E=0$ since we are now
considering the case in which  $\chi$ is independent of $\varphi$). 
The functions $A(r)$ to $D(r)$ are now of the form
\bear
A(r)&=&\Big[\,(n^2+1)(r-r_*)\Big]^{{3\over 2}}\,\,
\Big(\,1+o(r-r_*)\Big)\,\,,\rc\rc
B(r)&=&{(n^2+1)^{{3\over 2}}\over 2}\,\sqrt{r-r_*}\,\,
\Big(\,1+o(r-r_*)\Big)\,\,,\rc\rc
C(r)&=&{1\over 4}\,{(n^2+1)^{{3\over 2}}\over\sqrt{r-r_*}}\,\,
\Big(\,1+o(r-r_*)\Big)\,\,,\rc\rc
D(r)&=&\Big(r_*+{n^2-1\over 4}\,\Big)\sqrt{n^2+1}\,\sqrt{r-r_*}\,\,
\Big(\,1+o(r-r_*)\Big)\,\,.
\eear
Notice that, also in this case, the coefficients of the functions above are
such that, after a partial integration, the singular term of $C(r)$ cancels
against the leading term of $B(r)$. The differential equation that follows for 
$\xi$ in the IR is:
\beq
\partial_r^2\xi\,+\,\Big({3\over 2}\,{1\over r-r_*}\,+\,o
(r-r_*)\,\Big)\,
\partial_r\xi\,+\,o({1\over r-r_*})\,\xi\,=\,0\,\,,
\eeq
and, therefore, the indicial equation is the same as for eq. (\ref{IReq}) (\ie\
eq. (\ref{IRindicial})). It follows that also in this case there exists an
independent solution which does not diverge when $r\rightarrow r_*$.

\end{document}